%% file: draft3.tex
\documentclass[trackchanges, twocolumn, preprint2]{aastex701}
\usepackage{amsmath}
\usepackage{graphicx}
\usepackage{color}
\usepackage{underscore}
\usepackage{adjustbox}
\usepackage{booktabs}
\hypersetup{
    colorlinks,	
    citecolor=blue,
    }

\newcommand{\be}{\begin{eqnarray}}
\newcommand{\ee}{\end{eqnarray}}

\newcommand{\src}{Cyg~X-1}

\newcommand{\nustar}{{\it NuSTAR}}

\newcommand{\suzaku}{{\it Suzaku}}

\newcommand{\xrism}{{\it XRISM}}

\usepackage{dashrule}

\begin{document}

\title{Assessing systematic uncertainties from spectral re-analysis of Cyg X-1 with different coronal geometries}

\author[orcid=0009-0009-0287-2535]{Abdurakhmon~Nosirov}
\affiliation{Center for Astronomy and Astrophysics, Department of Physics, Fudan University, Shanghai 200438, China}
\email{abdurakhmonn24@m.fudan.edu.cn}

\author[orcid=0000-0002-9639-4352]{Jiachen~Jiang}
\affiliation{Department of Physics, University of Warwick, Coventry CV4 7AL, UK}
\email{jcjiang12@outlook.com}

\author[0000-0002-3180-9502]{Cosimo~Bambi}
\altaffiliation{bambi@fudan.edu.cn}
\affiliation{Center for Astronomy and Astrophysics, Department of Physics, Fudan University, Shanghai 200438, China}
\affiliation{School of Humanities and Natural Sciences, New Uzbekistan University, Tashkent 100001, Uzbekistan}
\email{bambi@fudan.edu.cn}

\author{John A. Tomsick}
\affiliation{Space Sciences Laboratory, 7 Gauss Way, University of California, Berkeley, CA 94720-7450, USA}
\email{jtomsick@ssl.berkeley.edu}

\begin{abstract}
In this work, we carry out a new spectral reanalysis of \nustar\ and \suzaku\ observations of the disk reflection spectra of the stellar-mass black hole X-ray binary \src~in the soft state. We compare three types of models: a broken power-law disk emissivity profile with no assumption about the coronal shape used in the previous work of the same observations, a compact lamppost corona, and an extended disk-like corona motivated by recent X-ray polarization results. Our goal is to measure the systematic uncertainties caused by the assumed geometry, with a focus on key parameters such as the black hole spin and the inclination of the inner accretion disk. We find that the disk-like corona gives a fit that is statistically similar to the broken power-law and lamppost models, but it leads to more physically reasonable results, such as a lower inclination angle of about $30^{\circ}$. By using a variable disk density model, we measure the disk density to be $n_{\rm e} \approx 10^{20}$\,cm$^{-3}$, which is similar to earlier results. While the extended corona model infers a wider allowed parameter space for black hole spin and the inner radius of the disk-shaped coronal region, this reflects the additional physical freedom of the model. Even so, the disk-like corona remains a strong and physically well-motivated candidate for explaining the X-ray emission from \src.
\end{abstract}

\section{Introduction}
\label{introduction}

Relativistic reflection features are commonly observed in the
X-ray spectra of black hole binaries \citep{Fabian:1989, Miller:2009, Fabian:2012, Miller:2013, Reynolds2014, Xu:2018, 2021ApJ...913...79T,jiang22} and active galactic nuclei \citep{Tanaka:1995, Nandra:2007, Brenneman:2011, Walton:2012, 2019ApJ...874..135T,jiang19c,jiang19b}.

It is now well established that the hard X-ray continuum emission originates from a ``corona'' of hot electrons which Compton-upscatters soft seed photons from the accretion disk \citep{Galeev:1979, Haardt:1993}. The geometry of this corona (its size, shape, location relative to the disk, and optical depth) is a key factor in determining several important observational features: the shape of the Comptonised continuum, the strength and shape of the reflected (disk-reprocessed) emission \citep{Wilkins:2011}, the time-lags (reverberation) between direct and reflected photons \citep{Wilkins:2016}, the degree and angle of polarisation of the emergent radiation \citep{Gianolli:2023}, and ultimately the energetics and coupling between disk, corona and (in some cases) jet \citep{Dauser:2013}. For example, different coronal geometries produce different illumination patterns of the disk, thereby affecting the ionisation structure, reflection fraction, and emissivity profile of the disk \citep{Fan:2025}. 

In the literature, one can identify several ``canonical'' coronal geometries~\citep{2021SSRv..217...65B}. These include:
\begin{itemize}
    \item A compact ``lamppost'' corona: a small region located on (or near) the spin axis of the black hole, irradiating the disk from above. This geometry produces steep disk emissivity profiles and strong relativistic effects in reflection \citep{Dauser:2013}.
    \item A slab (or sandwich) corona: a hot layer that covers the accretion disk (or part of it) as a slab or sandwich above/below the disk surface \citep{Wilkins2012, Poutanen:2017}. This geometry emphasises Comptonisation of seed photons from the disk and predicts significant soft-excess \citep{Dove:1997}.
    \item A spherical or ``sphere-plus-disk'' corona: a spherical or quasi-spherical hot plasma region that surrounds the inner disk or lies above it, with the disk truncated further out. This geometry often yields weaker reflection and different spectral shapes \citep{Dove:1997}.  
    \item A conical or funnel-shaped corona (e.g., due to the base of a jet), or a patchy/coronal cloud model \citep{Kim:2024}. Recent polarimetric and timing work suggests that simple lamppost geometry may be too restrictive \cite{Jana:2023}.
    \item  Cylindrical corona: a corona in the shape of a cylinder above/below accretion disk can be interpreted as the base of a jet \citep{Meng:2025}. This geometry can naturally explain weak disk reflection and strong hard X‐ray Comptonization.
\end{itemize} 

Choosing between these geometries affects not only the spectral fitting but also our view of the physics. For example, it helps determine whether the corona is compact and linked to the base of a jet \citep[e.g.,][]{wang21}, or extended and possibly part of the disk atmosphere; whether the disk is truncated or reaches the innermost stable circular orbit \citep[ISCO, e.g.,][]{Fabian:1989,basak17}; and whether the corona is moving or expanding, or instead remains static \citep[e.g.,][]{kara19}.

The target of this work, \src, is one of the brightest and best-studied black hole X-ray binaries in history and serves as an excellent test case for exploring different coronal geometries \citep[see a recent review in][]{Jiang:2024}. The system contains a stellar-mass black hole with a mass of about $20~M_\odot$ and a supergiant companion star \citep{Bolton1972, Webster1972, Orosz2011}. Because of its close distance and persistent X-ray emission, it is an ideal source for studying the inner accretion flow and the geometry of the surrounding corona. Over the years, many studies have modeled its X-ray spectrum, often focusing on the reflection features from the accretion disk and the power-law component produced by the corona. Disk reflection spectroscopy has been widely used to study the inner disk properties and geometry in this source \citep[e.g.,][]{Fabian:2012, 2019PhRvD..99l3007L}.

The most recent high-resolution X-ray spectra from the micro-calorimeters on \xrism\ reveal many spectral details and further confirm the existence of reflection spectra from the inner accretion disk of \src\ \citep{draghis25}. The first analysis of the \xrism\ data used a phenomenological broken power-law disk emissivity profile, without assuming a specific coronal geometry, found a high black hole spin of $a_{*}>0.98$ and a high inner disk inclination angle of $63^{\circ}$ \citep{draghis25}, which is significantly higher than the inclination angle of the binary orbit \citep{millerjones21}. Similar results were also reported in previous analyses \citep[e.g.,][]{Tomsick2014,Zhang:2020}. This work focuses on the observations taken during the soft/thermal emission-dominated state of \src, which were analyzed in \citet{Tomsick2014}.

Regarding the size of the coronal region in \src, earlier studies favored a compact corona \citep{Dove:1997, Tomsick2014, Tomsick2018}. The very steep disk emissivity profile suggests that the corona is confined within about 10 $r_{\rm g}$ \citep{Fabian:2012}. While the corona appears to be compact, its exact shape is still not well known. Some earlier works applied the lamppost model to the spectra of \src\ \citep[e.g., in the hard state,][]{parker15}. More recent polarization measurements in the hard state show that the polarization angle is aligned with the large-scale radio jet and suggest that a corona extended in the disk plane may better explain the polarization data \citep{Henric2022, Kravtsov:2025}. The measured high degree of polarization in \src\ exceeds the expected value for a compact corona. This points to a corona that is more extended than what is assumed in a simple lamppost model. Such an extended coronal geometry would lead to systematic changes in the inferred model parameters when fitting the observed spectral features. Additionally, it necessitates an inclination greater than the binary inclination, implying that the inner disk should be misaligned with respect to the binary orbital plane \citep{Henric2022, Krawczynski:2025}.

In this work, we study how different assumptions about the coronal geometry affect the spectral fitting of \src. We focus on the disk-like corona model developed by \citet{Shafqat2022} and compare its fit with both the conventional lamppost model and the broken power-law disk emissivity model. Our goals are: $i)$ to reproduce previous results obtained with the lamppost and broken power-law models, and $ii)$ to quantify the systematic uncertainties that arise from assuming different coronal geometries and to evaluate how these assumptions affect key spectral parameters, such as the black hole spin and the inclination of the inner disk. The primary aim of this work is not to determine the geometry of the coronal region in the soft state. Instead, we quantify the systematic uncertainties in reflection-based parameters (e.g., inclination, spin, emissivity profile) that arise from adopting different geometric prescriptions for the illuminating source.

This paper is organized as follows. In Section~\ref{data reduction}, we describe the observational data used in this study. In Section \ref{sec:extended_corona_model}, we outline the disk-like corona model and how it is used in our spectral fitting. In Section \ref{sec:fit}, we present the fitting results for the different coronal geometries and compare the best-fit parameters and fit quality for each model. In Section \ref{sec:emiss_profiles}, we show the uncertainties in the emissivity profiles of the best-fit models and discuss low inclination angle measured in this study. Finally, in Section \ref{sec:conclusion}, we summarize our main findings.

\section{Observations and Data Reduction}
\label{data reduction}

 \src\ was observed with {\em NuSTAR} and {\em Suzaku} on 2012 October 31 and November 1 (MJD 56,231 and 56,232). We analyze the same spectral data of the black hole binary Cyg X-1 that was used in \cite{Tomsick2014}. Although we present some important information about observations and the data reduction process, for detailed information we refer to \cite{Tomsick2014}. 

\subsection{NuSTAR}
{\em NuSTAR} data (ObsIDs 30001011002 and 30001011003) were reduced for Focal Plane Modules (FPMs) A and B, and each instrument has a total of approximately 15.5 ks exposure time. The data were reduced using version 1.1.1 of the NuSTARDAS pipeline software, with the 2013 May 9 version of the {\em NuSTAR} Calibration Database (CALDB)\footnote{We utilized the old pipeline software and CALDB to facilitate a direct comparison with the results of \citep{Tomsick2014}. We verified that data reduced using the latest pipeline version (v2.1.4) and the current \emph{NuSTAR} calibration files (as of December 15, 2025) yielded consistent results across all parameter values, with the sole exception of the cross-normalization constants.}. Event lists were cleaned using the \texttt{nupipeline} tool, and light curves and spectra were extracted using \texttt{nuproducts}. The source extraction region had a radius of $200^{\prime\prime}$, and the background region was chosen from a $90^{\prime\prime}$ circle located far from the source.

\subsection{Suzaku}

The {\em Suzaku} observation (ObsID 407072010) was reduced using the standard procedure. Data from the X-ray Imaging Spectrometers (XIS0 and XIS1)  cover the 0.3–10 keV energy range, and each has approximately 2 ks exposure time, while XIS2 was not operational during the observation and data from XIS3 were not used due to complications in its data reduction. As far as the HXD/PIN and HXD/GSO detectors are concerned, the PIN detector covers the energy range of 10–70 keV and has 30 ks exposure time, while the GSO detector extends the coverage to higher energies (60–600 keV) and has roughly 27.8 ks exposure.

\section{Extended corona model}
\label{sec:extended_corona_model}

\begin{figure}[htbp]
    \includegraphics[width=1\linewidth]{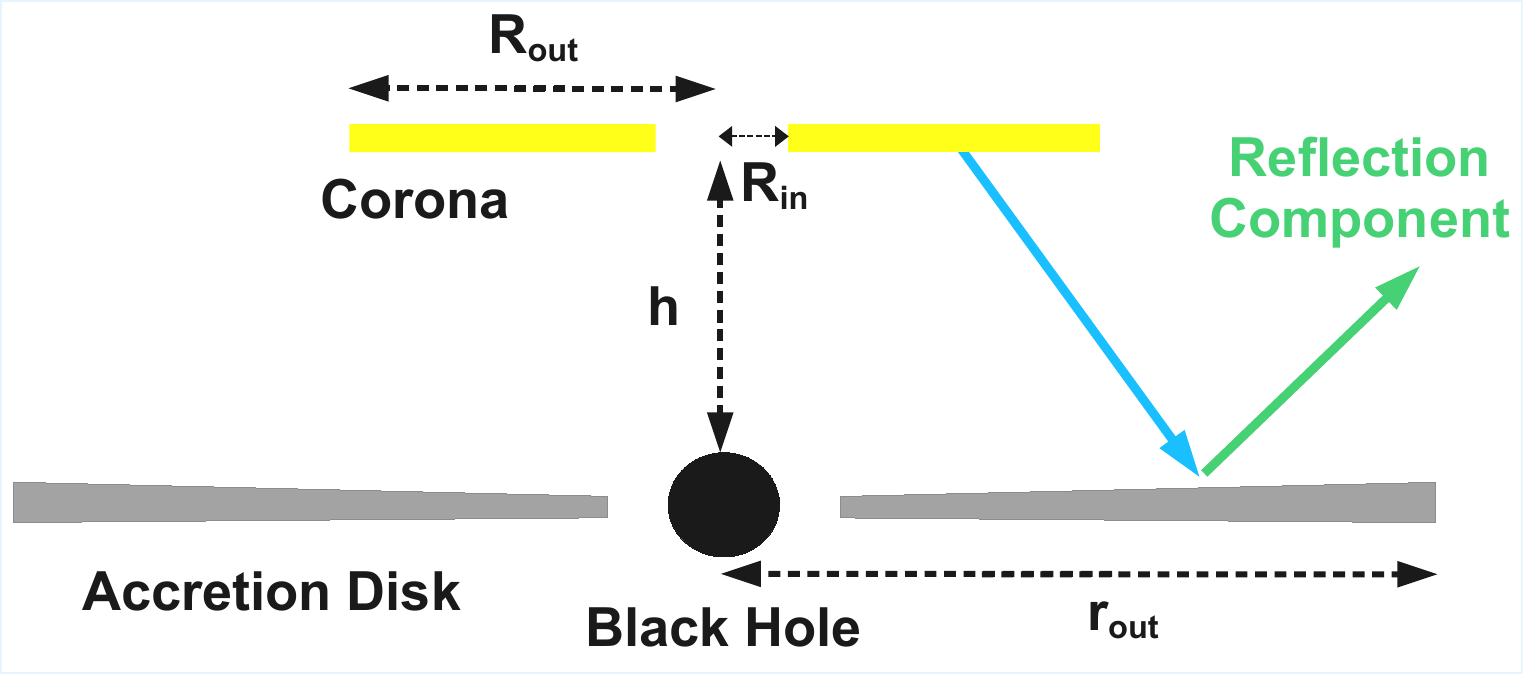}
    \caption{Cartoon of the astrophysical system. The corona is described by an infinitesimal thin disk of inner radius of $R_{in} $, outer radius of $R_{out} $ at height $h$ above the equatorial plane.}
    \label{fig:coronae}
\end{figure}

Disk-like or ring-like coronae have been discussed previously by \citet{Miniutti2003, Suebsuwong2006}. Several studies have computed the theoretical emissivity profiles of the accretion disk caused by irradiation from these coronal geometries, assuming Kerr spacetime \citep[e.g.,][]{Wilkins2012, gonzalez17, baker25}. A full reflection model, which convolves these profiles with a rest-frame reflection spectrum, was developed by \citet{Shafqat2022} for disk-like and ring-like coronae and is ready for data analysis. In this model, the corona is assumed to have a disk shape with infinitesimal thickness, located parallel to the accretion disk at a certain distance, referred to as the ``height'' ($h$) of the corona. An illustration of this geometry is shown in Fig.~\ref{fig:coronae}.

The corona has a central hole with radius $R_{\rm in}$. In the model, $R_{\rm in}$ is a free parameter with a lower limit of 0.5~$r_g$, where $r_g = M$ is the gravitational radius of the black hole. This central gap is included for two computational reasons: (1) rays starting from $R_{\rm in} < 0.5~r_g$ take much longer to reach the accretion disk, and (2) most photons emitted from $R_{\rm in} < 0.5~r_g$ fall into the black hole, contributing negligibly to the emissivity profile. The outer radius of the corona, $R_{\rm out}$, is also a free parameter with an upper limit of 24~$r_g$.

The disk-like corona consists of point-like sources that emit isotropically, meaning equal power is emitted into equal solid angles in the rest frame of each source. The point sources are placed at equal radial intervals ($\Delta R = \text{const}$), independent of their distance from the center. This setup is equivalent to assuming that the coronal intensity scales as $1/R$. Each photon’s path is traced numerically, including all relativistic effects, until it either hits the disk or falls into the black hole. Photons that land on the disk are radially binned over the disk surface. This process is repeated for every point source until $R_{\rm out}$ is reached\footnote{In the ray-tracing calculations of \citet{Shafqat2022}, intermediate processes, such as photons re-interacting and scattering within the corona due to strong gravitational bending, and Compton scattering of disk-reflected photons in the corona are ignored. Additionally, a stationary corona and a corona corotating with the disk below were compared, showing negligible differences \citep[see Fig.~3 in][]{Shafqat2022}. Therefore, the model in \citet{Shafqat2022} assumes a stationary disk-shaped corona.}. Further details can be found in \citet{Shafqat2022} and \citet{2024arXiv240812262B}. Figure~\ref{fig:coronae} is  schematic illustration of the extended disk-like coronal geometry used to compute the illumination pattern. Such a model can be interpreted as an effective description of emission from a vertically thin layer above the disk (like disk atmosphere/skin). This representation is phenomenological and does not imply a physically evacuated region between the disk and the corona.

\section{Fit with different coronal geometries}
\label{sec:fit}

This study aims to quantify the systematic uncertainties in spectral parameters that arise from the assumed coronal geometries. As a result, our analysis focuses on the coronal and reflection components of the spectrum. We follow the data analysis method of \citet{Tomsick2014}. Each subsection describes a specific model, its physical motivation, and the corresponding best-fit results. Our approach is to change one model component at a time to isolate and measure its effect on the spectral parameters and the goodness of fit. The full set of models is listed in Table~\ref{tab:models}.

We note that continuum-fitting methods (e.g., kerrbb/kerrbb2) can in principle provide complementary constraints on black hole spin, but require assumptions about the black hole mass, distance, and spectral hardening factor. Since our goal here is to isolate systematic effects associated with the reflection geometry and keep consistency with the previous works (e.g. \citet{Tomsick2014}), we adopt a phenomenological thermal component and defer a joint continuum–reflection analysis to future work. We used XSPEC 12.13.0c \citep{arnaud96} for spectral fitting, and the best-fit model parameters are determined by minimising the \(\chi^2\) statistic.

\input{t-models}

\subsection{Power-law emissivity profile}
\subsubsection{Model 0}

The earlier spectral analysis of these data was carried out by \citet{Tomsick2014}, who obtained the best fit using a phenomenological broken power-law emissivity profile with the \texttt{relconv} \citep{dauser10,dauser13} convolution model (labeled as Model~0 in Table~\ref{tab:models}). This setup uses the \texttt{reflionx_hc} reflection table, which was generated with the \texttt{reflionx} code \citep{Ross2005} assuming a fixed electron density of $10^{15}$\,cm$^{-3}$. We refer to this configuration as Model~0, as it serves as our baseline for reproducing the results of \citet{Tomsick2014} (where it was labeled ``Model~4''). Consistent with the \texttt{reflionx_hc} table, which assumes a \texttt{cutoffpl} incident spectrum, we use the same continuum model for the coronal emission component in Model~0. The inner radius of accretion disk \(r_{\mathrm{in}}\)\footnote{We used lowercase $r$ ($r_{\rm in}$ and $r_{\rm out}$) for accretion disk radius and uppercase $R$ ($R_{\rm in}$ and $R_{\rm out}$) for coronal disk radius (see Fig.~\ref{fig:coronae}).} is fixed at the ISCO, the outer radius \(r_{\mathrm{out}}\) is set to be \(400~r_g\), and the redshift is fixed at \(z = 0\). 

For modelling the absorption effects caused by the wind, we constructed a grid of table models utilizing XSTAR version 2.2.1bg \citet{Kallman2001}. We assumed solar abundances for all elements, fixed the number density at \(n = 10^{12} \, \text{cm}^{-3}\), and set the turbulent gas velocity at \(v_{\text{turb}} = 300 \, \text{km s}^{-1}\) \citet{Miller2002, Hanke2008}. 

\begin{figure}[htbp]
    \hspace{-1cm}
    \includegraphics[width=1.2\linewidth]{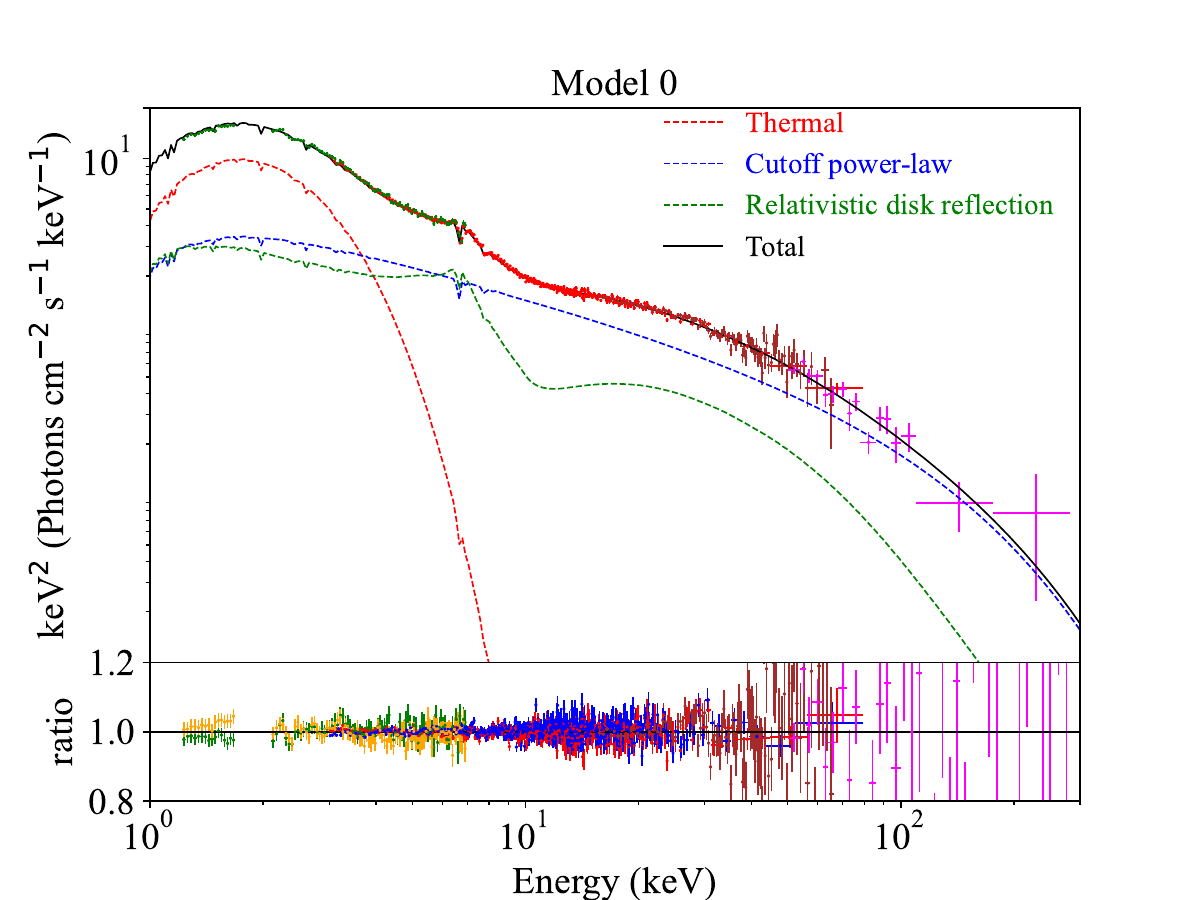}
    \caption{In the upper panel, the black solid line corresponds to the total model and other lines correspond to its components. In the lower panel, ratio plots of Model 0 are given and red, blue, green, orange, brown, and magenta correspond to {\em NuSTAR}'s FPMA, FPMB and {\em Suzaku}'s XIS0, XIS1, PIN, GSO data, respectively.}
    \label{fig:model0}
\end{figure}

We successfully reproduced the best-fit results of \citet{Tomsick2014} using Model~0. All best-fit parameters are listed in the first column of Table~\ref{relconv2} and are consistent with the values reported in \citet{Tomsick2014}. In Fig.~\ref{fig:model0}, we show each component of Model~0 along with the corresponding data/model ratio plots for illustration.

\input{t-relconv}

Analysis of the best-fit parameters in Table~\ref{relconv2} reveals a negative value for the outer emissivity index ($q_{\mathrm{out}}$). A negative outer emissivity index implies that emissivity increases with radius in the outer accretion disk ($r > r_{\mathrm{break}}$), which is unphysical because it would lead to a divergent disk luminosity. This suggests that additional illumination is required in the outer disk regions, potentially indicating a distant reflector, an extended coronal geometry, or a combination of both. 

To further validate our results, we examined the parameter constraints for all model combinations using the Markov Chain Monte Carlo (MCMC) algorithm. Specifically, we employed 500 walkers with a total chain length of 6,000,000 steps, discarding the first 1,000,000 steps as burn-in. The results for key parameters are presented in Appendix~\ref{sec:mcmc-plots} for some models, with contours corresponding to 68\%, 95\%, and 99.7\% confidence intervals.

To test this hypothesis, we first performed a subsequent fit with the outer index fixed to the theoretically expected value of $q_{\mathrm{out}} = 3$ for a flat spacetime. The resulting parameter values are presented in the second column of Table~\ref{relconv2}. Comparison of the $\chi^2$ values ($\Delta \chi^{2}=137.4$) shows that the fixed positive outer index $q_{\mathrm{out}} = 3$ is indeed statistically disfavored.

\subsubsection{Model 1}
\label{model1}

We then consider the addition of a distant reflection component to the model as an attempt to explain the negative outer emissivity index in Model 0. This is motivated by previous modelling of other observations of \src\ \citep[e.g.,][]{parker15,walton16,Tomsick2018}. More recent high-resolution X-ray spectra of \src\ obtained with \xrism\ further support the presence of narrow Fe K$\alpha$ doublet emission lines in the spectrum \citep{draghis25}. The narrow width of this line indicates an origin far from the black hole, likely at radii beyond $\sim1000\,R_{g}$ \citep{draghis25}, and provides a good fit to the narrow features observed in the \xrism\ data. The corresponding best-fit parameters of this model including a distant reflection component (Model~1) are presented in the third column of Table~\ref{relconv2}.

Model~1 provides a statistically significant improvement in the fit quality, with a reduction in the $\chi^2$ statistic of approximately 100 for only two additional free parameters compared to Model~0 with $q_{\mathrm{out}}=3$. However, the best-fit solution suggests a notably high ionization parameter of $\log\xi \approx 4.1$ for the distant reflector\footnote{It is worth noting that the inclusion of a distant reflector with fixed ionization at low values to Model~0, with \(q_{\mathrm{out}} = 3\), did not improve the statistics or alter the parameter values.}, which is unexpected if this material is located far from the black hole, as implied by the narrow width of the emission line. For comparison, the best-fit model presented in \citet{draghis25} used the \texttt{mytorus} model to describe this component, corresponding to reflection from neutral, cold gas. This discrepancy therefore warrants careful consideration.

\begin{figure*}[htbp]
    \centering
    \includegraphics[width=1.1\textwidth]{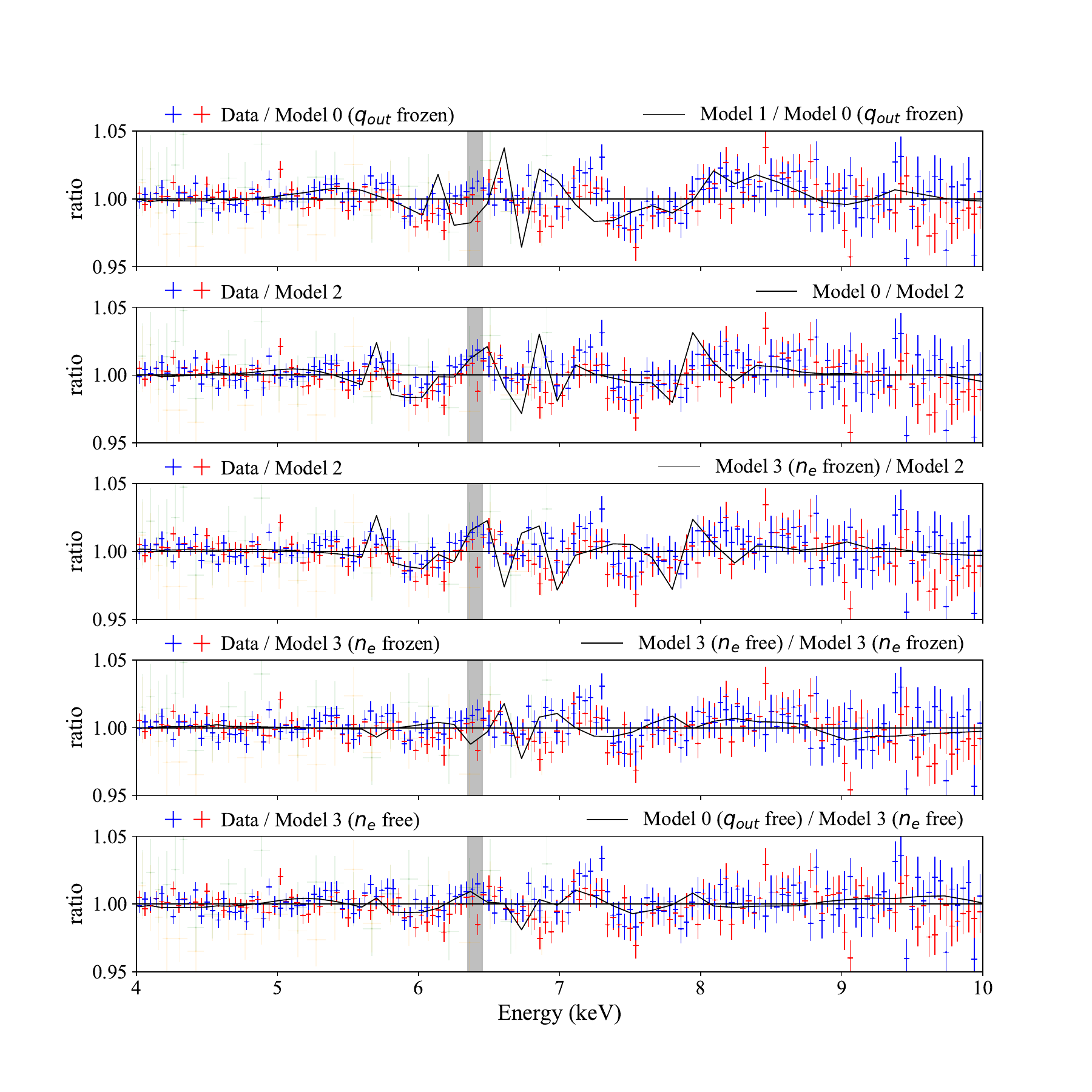}
    \caption{Error bars represent the ratio plots (data/model) for the models in the denominator for each case (see upper left corner of each plot). The black broken solid lines show the ratios of different models, as indicated in the upper right corner of each plot. This allows us to clearly assess how each model in the numerator improves the fit of the model in the denominator. Since significant changes are observed primarily in the 4-10 keV energy range, the ratios are displayed within this range. Ratio plots with whole energy range (1-200 keV) for all model combinations are presented in Fig.~\ref{fig:ratio-full}. Red, blue, faint green, and faint orange correspond to {\em NuSTAR}'s FPMA/FPMB and {\em Suzaku}'s XIS0/XIS1 data, respectively. The thick gray line is centered at the iron K$\alpha$ line (6.4 keV), with a width of 0.1 keV.}
    \label{fig:ratios}
\end{figure*}

An examination of the residual structure, shown in the first panel of Fig.~\ref{fig:ratios} (Model~1 / Model~0 with frozen $q_{\mathrm{out}}$), focuses on the 4--10\,keV iron emission band. The data points represent the Data/Model~0 ratio, while the solid lines show the Model~1/Model~0 ratio. Agreement between the two therefore highlights the energy ranges where Model~1 improves upon Model~0. The remaining panels in Fig.~\ref{fig:ratios} illustrate analogous comparisons for the other model pairs. From this analysis, we find that Model~1 provides an improved description of the spectrum in the 7.5–9\,keV energy range where the blue wing of the broad Fe K emission line is. However, the improvement around the Fe K$\alpha$ emission line between 6--7~keV remains limited. Together with the very high inferred ionization parameter, this indicates that a distant reflection component alone cannot fully account for the requirement of a negative outer disk emissivity index in Model~0. In the subsequent analysis, we therefore consider the effects of an extended coronal geometry, a lower system inclination, and a combination of both, in order to further improve the fit in the iron emission band.

\subsection{Horizontally extended corona}
\label{sec:extended_corona}

Motivation for considering a horizontally extended coronal geometry arises from recent polarimetric observations of this source \citep[e.g.,][]{Henric2022}. To test this scenario spectroscopically, we employed the disk-like corona model \texttt{relconvdisk\_nk} \citep{Shafqat2022} introduced in Sec.~\ref{sec:extended_corona_model}.

\subsubsection{Model 2}
\label{model2}

Model 2 replaces the \texttt{relconv} convolution kernel in Model 0 and Model 1 with \texttt{relconvdisk\_nk} and substitutes the reflection table with \texttt{reflionx\_HD\_nthcomp\_v2\_log} \citep{Jiachen2020} also generated using the \texttt{reflionx} codes \citep{Ross2005}. A motivation to change the reflection table was to utilise a reflection model where the electron density is a free parameter, allowing us to investigate its effects on the spectral fit\footnote{For comparison with Model 0, we also performed fits using \texttt{relconv} together with \texttt{reflionx\_HD\_nthcomp\_v2\_log} table by fixing the electron density at $10^{15}$ cm$^{-3}$, which yielded consistent results.}. Since the \texttt{reflionx\_HD\_nthcomp\_v2\_log} table assumes \texttt{nthComp} as the incident spectrum, we adopted this model for the Comptonized continuum component in our spectral fitting. The best-fit parameters for Model 2 are presented in the first column of Table~\ref{tab:relconvdisk}. In our fits, the coronal height parameter is constrained to be small of a few gravitational radii (see Model 3 ($n_e$ frozen) and Model 3 ($n_e$ free)), such that the geometry can be interpreted as an effective vertically thin emitting layer above the disk rather than a detached structure separated by empty space. We do not model the vertical stratification of the disk atmosphere explicitly; the geometry is intended as a tractable approximation motivated by recent polarimetric results indicating extended emission parallel to the disk plane.

\input{t-relconvdisk}

When using Model~2 without a distant reflection component and with the disk density fixed at $10^{15}$\,cm$^{-3}$, we find that the best-fit inclination angle is significantly lower ($i = 27.0^{+1.4}_{-0.5}$\,$^{\circ}$) than in the broken power-law emissivity profile cases (Models~0 and~1). This value is consistent with the binary orbital inclination of $i = 27.1 \pm 0.8^{\circ}$ \citep{Orosz2011}. Model~2 yields a lower limit on the iron abundance (Fe/solar $> 4.94$) and on the coronal height ($h > 6.7$). The black hole spin is constrained to be $a_{*} = 0.82^{+0.08}_{-0.38}$ (see Table~\ref{tab:relconvdisk}).\footnote{The outer radius ($R_{\mathrm{out}}$) of the disk-like corona was fixed at 24\,$r_g$ in all analyses due to model uncertainties at larger coronal radii. This value represents the maximum radius for which the model calculations are considered reliable. Although freeing this parameter yields a marginal improvement in fit quality ($\Delta\chi^2 \approx 7$), it has a negligible impact on the other spectral parameters.}

Compared to Model~0 with a fixed outer emissivity index, Model~2 provides a statistically significant improvement in fit quality ($\Delta\chi^{2} = -63.4$ for the same number of free parameters). However, it still yields a worse fit than Model~0 with a variable, negative best-fit outer emissivity index ($\Delta\chi^{2} = 74$ with one fewer free parameter). A comparison of the residuals between Model~0 with variable emissivity and Model~2 (second panel of Fig.~\ref{fig:ratios}) shows that the higher fit quality of Model~0 is primarily driven by its improved characterisation of the iron K$\alpha$ emission line around 6.4\,keV.

\subsubsection{Model 3 ($n_e$ frozen)}

To further improve the fit of the 6.4\,keV iron line based on Model~2, we incorporated an additional distant reflection component and refer to this configuration as Model~3 (see Table~\ref{tab:models}). The fitting results for Model~3 with the electron density fixed ($n_{\rm e}$ frozen) are presented in the second column of Table~\ref{tab:relconvdisk}. The inclusion of the distant reflector improves the fit by $\Delta\chi^{2} = 51.9$ for two additional free parameters. Notably, the ionisation state of the distant reflector is now more physically reasonable ($\log\xi \approx 2.2$), in contrast to the high ionisation inferred in Model~1. Compared to Model~2, the inferred iron abundance is reduced (Fe/solar $\approx 3.8$), and the coronal height becomes constrained from above ($h < 5.3$).

The changes in the iron abundance and coronal size arise because, in Model~3, the distant reflection component accounts for the narrow Fe K$\alpha$ emission line, whereas in Model~2—where no distant reflector is included—the narrow line is instead reproduced by invoking a more extended coronal region, resulting in an artificially narrower model line profile. A direct comparison between Model~2 and Model~3 with fixed $n_{\rm e}$, shown in the third panel of Fig.~\ref{fig:ratios} (Model~3 / Model~2), clearly demonstrates that Model~3 provides a better description of the data, particularly in the iron K$\alpha$ line region around 6.4$\,$keV.

\subsubsection{Model 3 ($n_e$ free)}

In all previous analyses, the electron density was fixed at $10^{15}$\,cm$^{-3}$. When this parameter is allowed to vary, motivated by the high disk density inferred for \src\ in \citet{Tomsick2018}, we observe only a marginal improvement in fit quality ($\Delta\chi^{2} \approx 11$), primarily driven by a better description of otherwise unstructured residuals in the spectrum (see the fourth panel of Fig.~\ref{fig:ratios}, Model~3 with $n_{\rm e}$ free / Model~3 with $n_{\rm e}$ frozen). However, freeing the electron density leads to significant changes in several key spectral parameters. Specifically, the inferred iron abundance decreases further (Fe/solar $\approx 2$), and the ionization parameter of the main reflector shifts to substantially lower values, consistent with the results of \citet{Tomsick2018}.

In Fig.~\ref{fig:model3-final}, we show the individual spectral components of Model~3 together with the corresponding ratio plots for illustrative purposes (see the figure caption for details). By comparing Fig.~\ref{fig:model0} and Fig.~\ref{fig:model3-final}, we find that the Compton component of the spectrum is notably weaker in the latter case. In other words, flux ratio between reflection and Compton component of spectrum ($F_{\rm reflection}/F_{\rm Compton}$) is $0.78$ and $2.26$ in Model 0 (with free $q_{out}$) and Model 3 (with free $n_e$), respectively.

\begin{figure}[htbp]
    \hspace{-1cm}
    \includegraphics[width=1.1\linewidth]{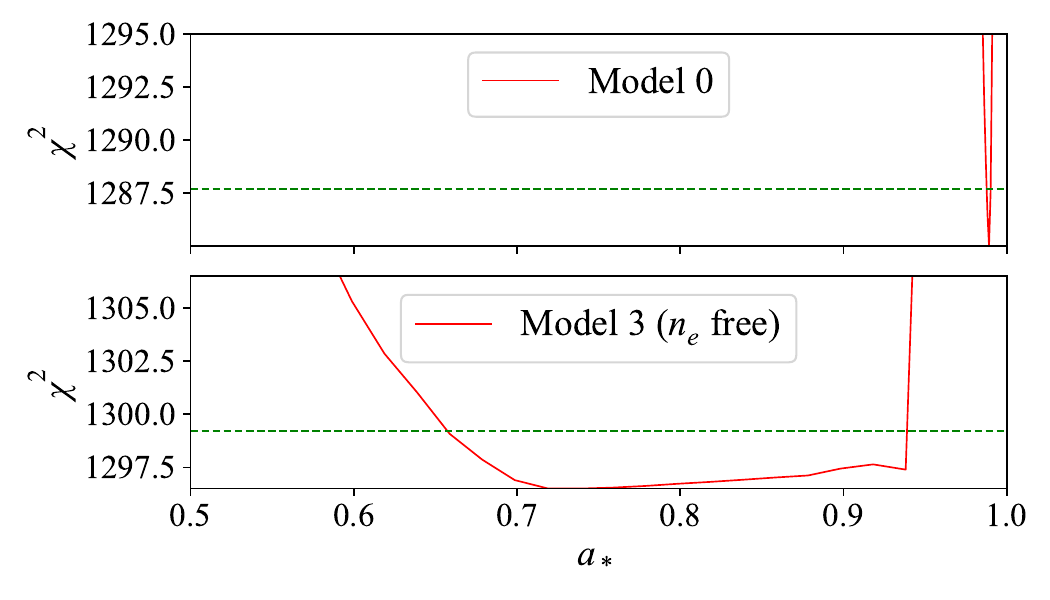}
    \caption{One-dimensional confidence intervals for the spin parameter across the different models considered in this work. The green dashed line indicates the 90\% confidence level ($\Delta \chi^2 = 2.706$).}
    \label{fig:spin}
\end{figure}

\begin{figure}[htbp]
    \hspace{-1cm}
    \includegraphics[width=1.2\linewidth]{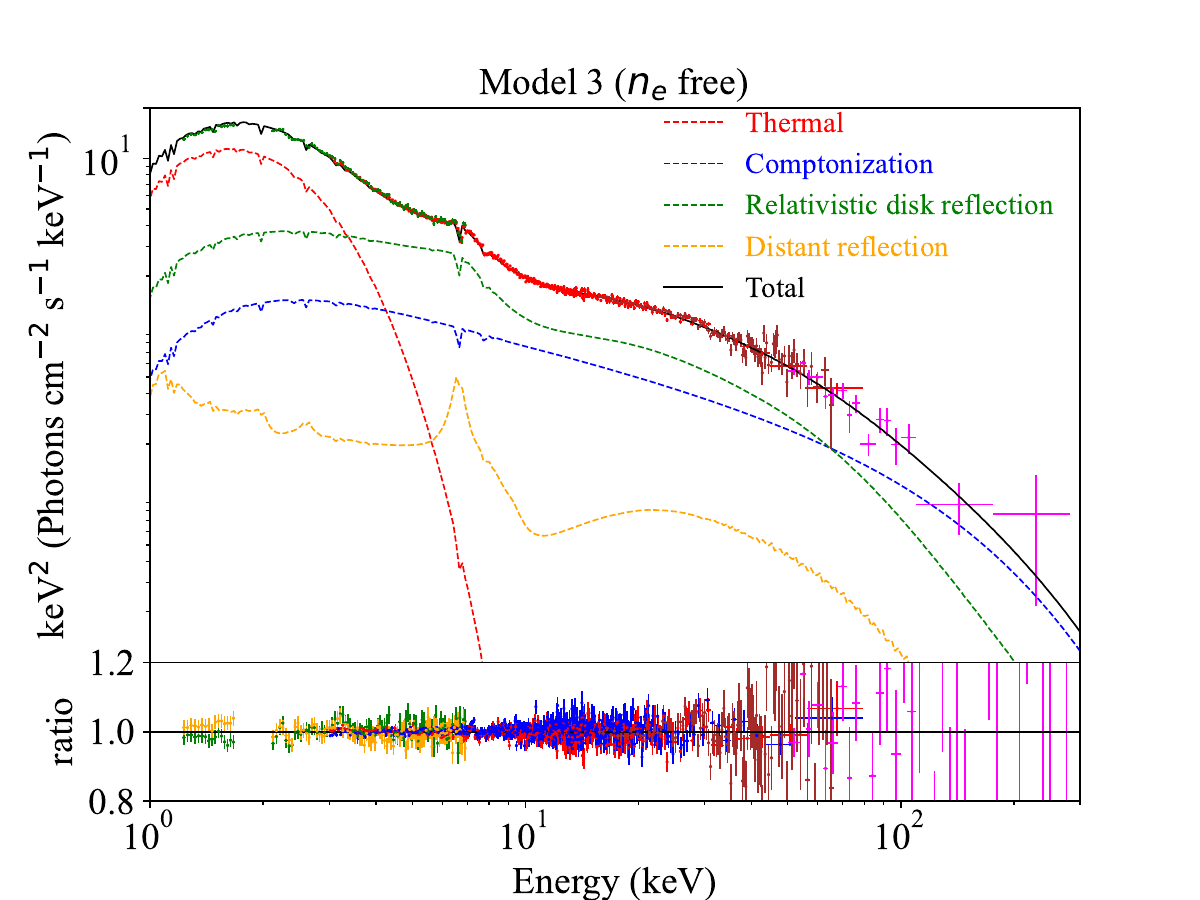}
    \caption{In the upper panel, the black solid line corresponds to the total model and other colourful lines correspond to its components. In the lower panel, ratio plots of Model 3 are given and red, blue, green, orange, brown, and magenta correspond to {\em NuSTAR}'s FPMA, FPMB and {\em Suzaku}'s XIS0, XIS1, PIN, and GSO data, respectively. See the text for more details.}
    \label{fig:model3-final}
\end{figure}

The electron density itself is now relatively well constrained at a value significantly higher than the previously assumed $n_{\rm e} = 10^{15}$\,cm$^{-3}$, converging to approximately $10^{20}$\,cm$^{-3}$ \citep{Jiang2019}, in line with values inferred for other accreting black hole systems \citep[e.g.,][]{jiang20}. The black hole spin parameter is constrained to $a_* = 0.73^{+0.21}_{-0.07}$ . Although the statistical uncertainty is larger than that obtained for the best-fit Model~0, it is nonetheless consistent within uncertainties (see the \(\chi^{2}\) distribution in Fig.~\ref{fig:spin} as a function of the spin parameter, which is calculated using the \texttt{steppar} function in \textsc{XSPEC}, and the MCMC results in Fig.~\ref{fig:mcmc-relconvdisk}).

As a final assessment of the disk-like coronal geometry for Cyg~X-1, we compare our best-fit model incorporating a free electron density (Model~3 with $n_{\rm e}$ free) with the best-fit model from \citet{Tomsick2014} (Model~0). The statistical difference between these two models is relatively small, with $\Delta\chi^{2} \approx 11$. This difference could be further reduced through additional parameter exploration, as discussed earlier, for example by freeing the outer radius of the disk-like corona ($R_{\rm out}$). The remaining discrepancy is primarily driven by Model~0 providing a slightly better description of the residual unstructured features in the spectrum (see the final panel of Fig.~\ref{fig:ratios}).

\subsection{Lamppost corona}
\label{sec:lamp_post}

At present, Model~0 provides the statistically best fit to the data; however, it requires an unphysical negative outer emissivity index. In contrast, Model~3 with a free electron density ($n_{\rm e}$ free) yields a slightly worse fit, with $\Delta\chi^{2} \approx 11$, but resolves the issue of the negative outer emissivity by invoking a combination of a horizontally extended coronal region and an additional distant reflection component. This geometry is also independently supported by X-ray polarization measurements of the source.

If polarimetric constraints were not taken into account, we would instead consider testing a more sophisticated relativistic reflection model, \texttt{relxilllpCp} \citep{dauser13}, which incorporates a lamppost coronal geometry\footnote{At the time of writing, a recent update to the \texttt{Relxill} model also incorporates a similar disk-shaped coronal geometry. A separate independent study is currently underway to compare our models with this newly updated version.} and provides a more detailed physical treatment of the reflection process.

We emphasize that the lamppost geometry employed in reflection models used for data analyses \citep{dauser13} should not be interpreted as a physical jet model. Rather, it provides a simplified geometric prescription for compact illumination located near the spin axis, introduced primarily to make relativistic ray-tracing calculations tractable. Lamppost-type models are also widely applied to radio-quiet AGN where no steady or significant jet is detected, demonstrating that such a geometry should be regarded as an effective description of compact coronal emission rather than as a jet-launching structure \citep{Jiang:2024}.

We have therefore explored two such configurations, referred to as Model~4 and Model~5, which share two key improvements in modelling in comparison to Model 3:

\begin{itemize}
\item Inclusion of returning radiation effects, accounting for photons that are reflected from the disk and subsequently re-illuminate the disk\footnote{We note that the treatment of returning radiation in the model involves several important simplifications, as discussed in \citet{2024ApJ...965...66M} and \citet{2024ApJ...976..229M}.}
\item Implementation of a power-law ionization profile across the accretion disk to approximate a more realistic disk, rather than assuming a constant ionization parameter.
\end{itemize}

\subsubsection{Model 4}
\label{model4}

Model~4 (see Table~\ref{tab:models}) includes a separately added Comptonisation component, with the reflection fraction fixed at $-1$ so that the \texttt{relxilllpCp} component returns only the disk reflection spectrum. In this way, the overall strength of the reflection component remains a free parameter, consistent with the treatment in Models~0–3. The corresponding best-fit parameters are listed in the first column of Table~\ref{tab:relxilllpCp2}, and the MCMC results for Model~4 are shown in Fig.~\ref{fig:mcmc-relxillCp}. A comparative analysis reveals several distinctive characteristics of this model:

\begin{itemize}
\item Statistically, this configuration provides the best fit among all models explored in this work, with $\chi^{2}/\nu = 1240.3/1093$.
\item The inferred inclination angle ($i \approx 40^{\circ}$) is higher than those obtained with disk-like coronal geometries (typically $\sim30^{\circ}$), but remains significantly lower than the value inferred from Model~0 ($i \approx 70^{\circ}$).

\item The black hole spin parameter is consistent with a relatively lower spin (compared to other studies of this source), with $a_* \approx 0.88$.

\item The photon index is lower ($\Gamma \approx 2.42$) than in all other models investigated (e.g., $\Gamma \approx 2.59$ in Model~0 and $\Gamma \approx 2.56$ in Model~3 with free electron density).

\item The column density of interstellar absorption is constrained to a higher value ($N_{\rm H} \approx 6.53 \times 10^{21}$\,cm$^{-2}$) compared to Model~3 with free electron density ($N_{\rm H} \approx 4.95 \times 10^{21}$\,cm$^{-2}$).

\item The electron density is well constrained but at lower values than those obtained in Model~3 with free electron density. 
\end{itemize}

\input{t-relxilllpCp}

\subsubsection{Model 5}

In Model~5, we adopt the fully self-consistent treatment of the reflection strength within the \texttt{relxilllpCp} framework, in which the reflection fraction is calculated directly from the black hole spin ($a_*$) and the coronal height ($h$), rather than being introduced as an independent parameter. In this configuration, the \texttt{relxilllpCp} component simultaneously returns both the primary Comptonised emission and the disk reflection spectrum. Notably, the resulting best-fit parameter values and statistical quality of the fit (see the second column of Table~\ref{tab:relxilllpCp2}) are very similar to those obtained for Model~4 (see Fig.~\ref{fig:mcmc-relxillCp-boost} for the corresponding MCMC results). This indicates that the observed strength of the disk reflection component can be naturally explained within the lamppost geometry scenario. Based on the spectral analysis alone, we are unable to statistically distinguish whether the lamppost geometry or the extended coronal geometry provides a superior description of the data.

\section{Discussions}
\label{sec:emiss_profiles}

\begin{figure}[htbp]
    \hspace{-1cm}
    \includegraphics[width=1.1\linewidth]{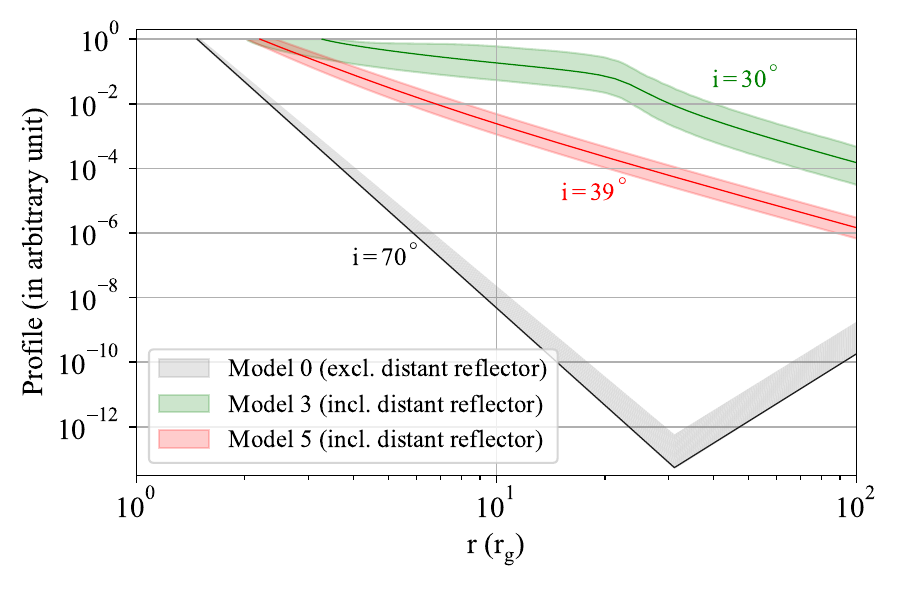}
    \caption{Emissivity profiles for Models: 0, 3, 4 and 5 together with shaded 90\% confidence intervals. See the text for more information.}
    \label{fig:emiss}
\end{figure}

In this section, we will further discuss uncertainties in the emissivity profiles of different model combinations.
The shaded regions in Fig.~\ref{fig:emiss} indicate the 90\% uncertainty ranges based on the 90\% confidence intervals constrained by each model, and solid lines are the best-fit emissivity profiles.

Model~0 exhibits the steepest emissivity profile, with relatively higher emission from the outer disk. This component is effectively replaced by the distant reflector in subsequent models. An extended coronal region produces an emissivity profile that is initially steep, then flattens, and subsequently decreases with a slope of \(q = -3\). The lamppost geometry provides the best statistical fit to the data, with a \(\Delta\chi^2\) improvement of 45 and one additional free parameter compared to Model~0. However, the extended coronal region is still physically favored, as it is more consistent with polarization measurements.

\subsection{Inclination angle}
We obtain a low inclination angle ($\sim30^\circ$) by assuming a horizontally extended corona and an additional distant reflector. This result is consistent with the known binary orbital inclination ($27.1^\circ \pm 0.8^\circ$; \citealt{Orosz2011}). Furthermore, recent radio jet analysis of this source by \citet{Prabu2025} yields consistent results with very small misalignment between the jet and the orbital axis.

In contrast, polarimetric measurements reported by \citet{Henric2022} suggest that a "sandwich" corona configuration—coupled with a standard accretion disk extending to the innermost stable circular orbit (ISCO)—requires an inclination exceeding $65^\circ$. Even considering a truncated disk, their model requires an inclination greater than $45^\circ$ to account for the observed polarization degree. These values conflict with our findings of $i \approx 30^\circ$ based on a disk-like corona geometry. Additionally, \textit{XRISM} data analysis by \citet{draghis25} revealed results similar to our Model 0, with an outer emissivity index of $q_{\text{out}} \approx 1.8$--$1.9$ and a high inclination of $i \approx 62^\circ$--$65^\circ$. However, a direct comparison is limited as the source was in the hard state during their observations.

Nevertheless, \citet{Poutanen2023} demonstrated that a low inclination of $\sim 30^\circ$ can be reconciled with a high polarization degree if the horizontally extended emitting plasma (e.g., corona) is outflowing at mildly relativistic velocities ($\beta \geq 0.4$; see Fig.~2 in \citeauthor{Poutanen2023}). Our current geometric configuration does not account for relativistic outflows or the Comptonization of soft photons by matter undergoing relativistic bulk motion (like the \texttt{bmc} model in \texttt{XSPEC}). To refine these constraints, future work should implement models featuring an outflowing extended corona and perform simultaneous spectro-polarimetric fitting. Such an approach will provide a more robust assessment of the system geometry by satisfying both spectral and polarization constraints concurrently.

\begin{figure*}
    \hspace{-0.5cm}
    \includegraphics[width=1.1\linewidth]{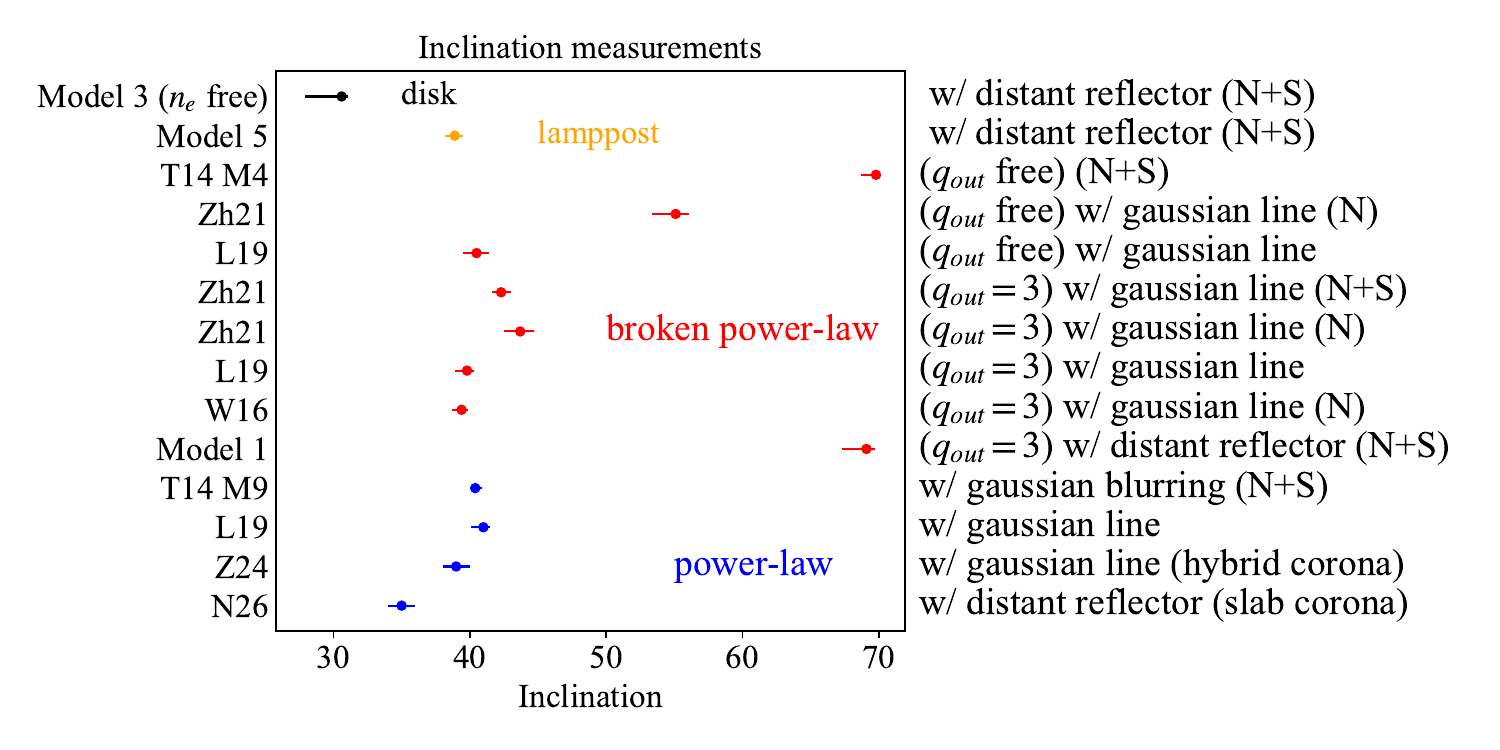}
    \caption{Comparison of inclination angle measurements for \src\ from the present study and previous works. The colors and text on the right side of the plot indicate the emissivity profiles used in each analysis. The labels $N$ and $N+S$ on the right-hand side of the figure denote studies based on the same NuSTAR-only observations as in this work, or on the combined NuSTAR+Suzaku dataset, respectively. Entries without these labels correspond to different soft-state observations of Cyg~X-1. The following abbreviations are used in the plot: T14 - \citet{Tomsick2014}, W16 - \citet{Walton2016}, L19 - \citet{Liu:2019}, Zh21 - \citet{Zhang:2020}, Z24 - \citet{Zdziarski:2024}, N26 - \citet{Niedzwiecki:2026}, where ``T14 M4'' and ``T14 M9'' refer to Model~4 and Model~9 of \citet{Tomsick2014}, respectively.}

    \label{fig:inclination-measurements}
\end{figure*}

Since the primary objective of this work is to quantify the systematic uncertainties associated with different coronal geometries, it is useful to compare our measurements of the key parameters with previous soft-state measurements of Cyg~X-1. Figure~\ref{fig:inclination-measurements} presents a comparison of the soft-state inclination angle measurements of \src\ reported in the literature, together with the results obtained in the present work. The data points are shown with their $90\%$ confidence intervals and are color-coded according to the adopted emissivity prescription, namely power-law, broken power-law (with either free or fixed outer index, $q_{\rm out}$), lamppost, and disk-like geometries.

In addition, different studies adopted different approaches to account for excess emission originating in the outer accretion disk, in particular the line feature around 6.4~keV. Some authors, including \citet{Zdziarski:2024} and this work, modeled this component using a distant reflector (indicated as ``w/ distant reflector'' on the right-hand side of the figure), whereas others employed an additional Gaussian line component \citep{walton16, Liu:2019, Zhang:2020, Niedzwiecki:2026} (indicated as ``w/ gaussian line''). In other cases, the outer emissivity index of the broken power-law profile was allowed to take negative values in order to mimic enhanced emission at larger radii, as in our Model~0, which reproduces Model~4 of \citet{Tomsick2014}. The labels $N$ and $N+S$ on the right-hand side of the figure indicate studies based on the same NuSTAR-only observations as used here, or on the same combined NuSTAR+Suzaku data set, respectively. Entries without these labels correspond to different soft-state observations of Cyg~X-1.

Figure~\ref{fig:inclination-measurements} clearly demonstrates that the inferred inclination angle depends on the assumed emissivity profile and the overall model composition. In particular, the inclusion of an additional Gaussian line component, either alone or in place of a distant reflector, in combination with phenomenological power-law or broken power-law emissivity profiles, tends to yield substantially lower inclination angles ($\sim 40^\circ$). By contrast, the inferred inclination rises to $\sim 70^\circ$ when the Gaussian component is omitted (Model~0) or when a distant reflector is used instead (Model~1). This behavior was already noted by \citet{Tomsick2014} (see their Models~4 and 9). A related example is provided by \citet{Zhang:2020}, who analyzed the same NuSTAR observation using broken power-law emissivity profiles with an additional Gaussian line, and obtained inclination angles of $43.7^{+1.0}_{-1.2}$ and $55.1^{+1.0}_{-1.7}$ for fixed ($q_{\rm out}=3$) and free outer emissivity indices, respectively (see their Table~VI).

When a certain coronal geometry is assumed in the model, the disk-like coronal geometry, for instance, exhibits different behaviors. This model infers a much lower inclination angle, $\sim 30^\circ$, with or without including an additional distant reflection component (Models 2 and 3). This value is considerably more consistent with the binary inclination. The statistical uncertainties are consistent across the different models, indicating that the dominant differences arise from model-dependent systematics rather than from formal measurement precision.

When a disk-shaped coronal geometry is assumed, the outer disk receives more illumination than in the case of a compact coronal geometry, such as the lamppost geometry. This enhanced illumination produces a stronger blue wing in the line profile, which in turn leads to a lower inferred inclination angle. In contrast, the very steep emissivity profile in the inner disk inferred from the best-fit Model 0 would require most of the emission to be highly redshifted due to gravitational redshift. Consequently, a much higher disk inclination angle is needed to fit the broad Fe K emission line in the data, and this ultimately yields the unphysical negative emissivity index in Model 0. Refer to Fig.~\ref{fig:emiss} for the emissivity profiles of the different best-fit models.

\subsection{Spin}

\begin{figure*}
    \hspace{-0.5cm}
    \includegraphics[width=1.1\linewidth]{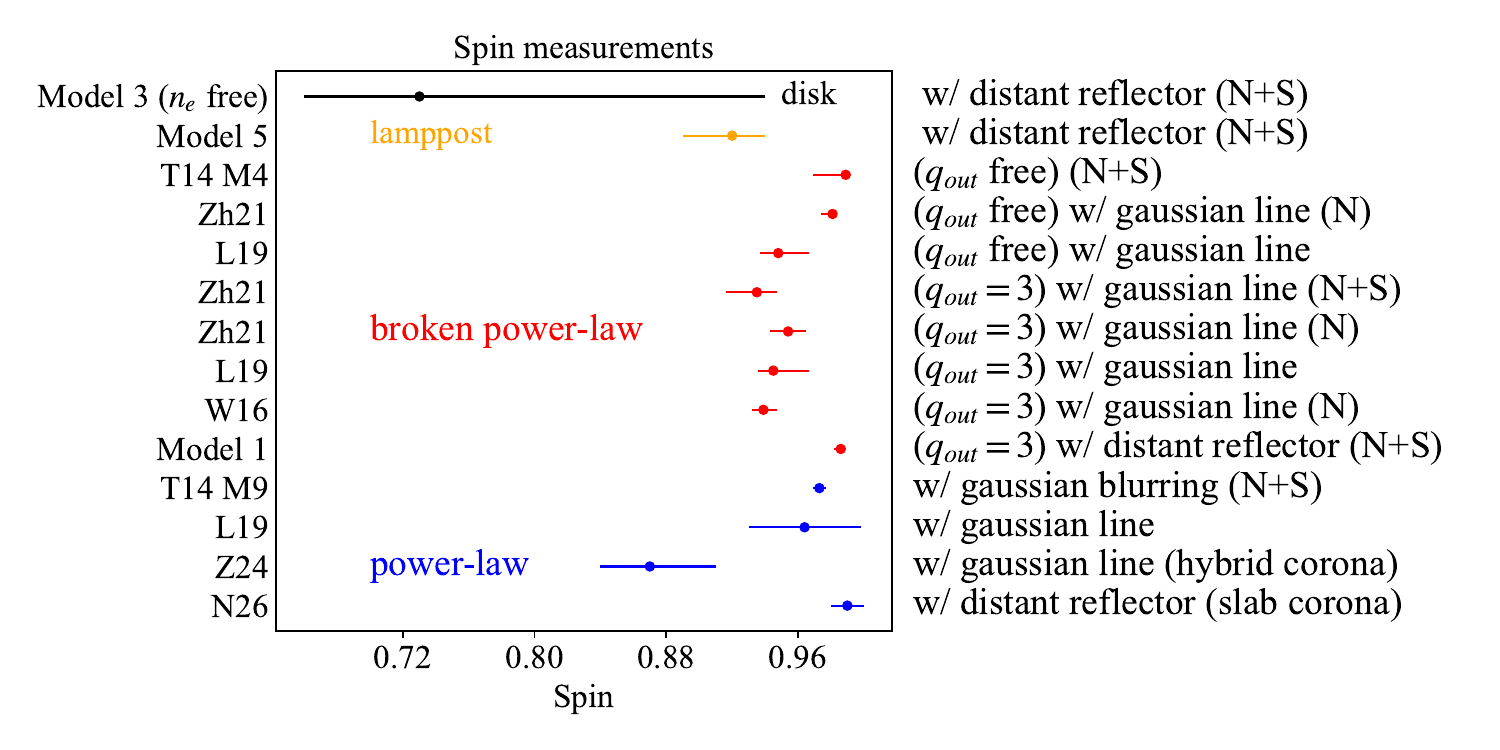}
    \caption{Comparison of spin measurements of the \src\ from this work and previous studies, with colors indicating the emissivity profiles adopted in each analysis.}
    \label{fig:spin-measurements}
\end{figure*}

Figure~\ref{fig:spin-measurements} shows a comparison of spin measurements from this work and previous studies. Statistical uncertainties are relatively small for power-law and broken power-law emissivity profiles, with the spin consistently constrained to high values. In the case of \citet{Zdziarski:2024}, authors applied convolution hybrid (thermal/non-thermal) hot plasma emission model (\texttt{ceqpair}) and \texttt{kerrbb} model with free color correction factor ($f_{\rm col}$) for the thermal emission. However, authors also showed that by fixing the inclination at the binary value ($i = 27.5^\circ$) and the color factor ($f_{\rm col} = 1.7$), they obtained a similarly tight constraint on the spin ($a = 0.986^{+0.002}_{-0.001}$) with comparable fit statistics.

In contrast, for the lamppost geometry (Model 5 in this study), the best-fit spin is slightly lower, and the associated statistical uncertainties are moderately larger. The effect is most pronounced in the disk-like geometry, where statistical uncertainties dominate, yielding significantly less precise spin constraints compared to the other emissivity profiles.

\subsection{Ionization and iron abundance}

\begin{figure*}
    \hspace{-0.5cm}
    \includegraphics[width=1.1\linewidth]{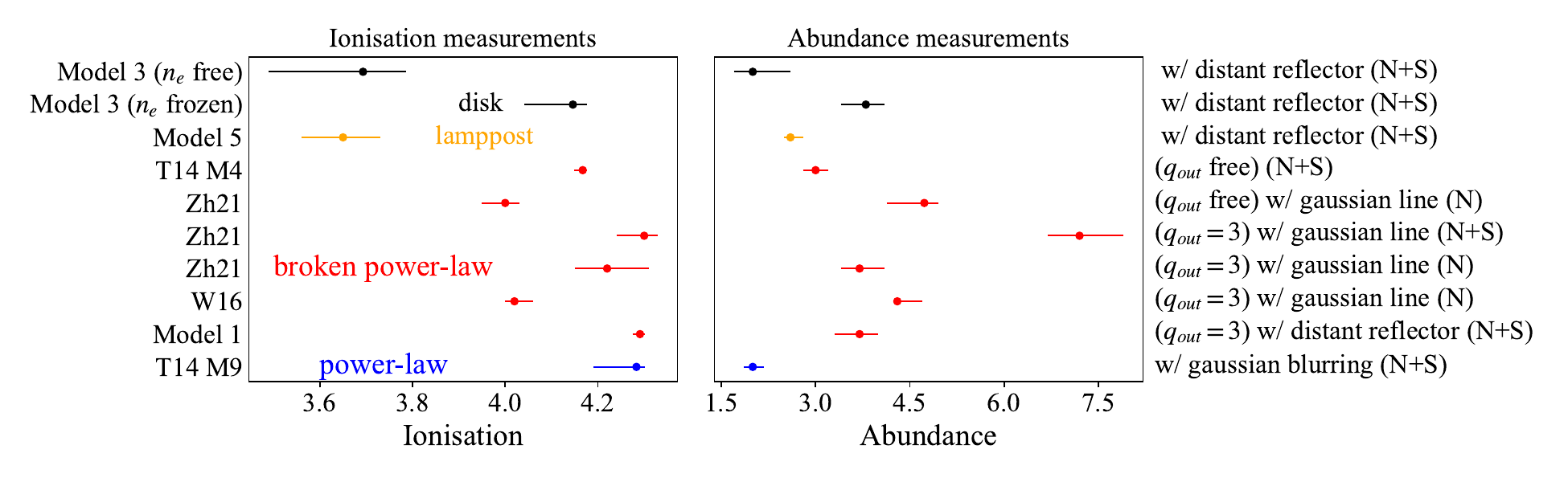}
    \caption{Comparison of ionization and iron abundance measurements of the same \textit{NuSTAR} or \textit{NuSTAR+Suzaku} observations of \src\ in different studies, with colors indicating the emissivity profiles adopted in each analysis.}
    \label{fig:ion-abund-measurements}
\end{figure*}

Figure~\ref{fig:ion-abund-measurements} compares the ionization and iron-abundance measurements reported for \src\ in studies analyzing the same \textit{NuSTAR} or \textit{NuSTAR+Suzaku} observations. The left panel presents the inferred ionization parameter. Consistent with its definition, $\xi = 4\pi F / n_e$, this comparison indicates that, within the set of models considered for the observations analyzed in this work, the inferred ionization value depends on the assumed accretion-disk electron density ($n_e$) than to the adopted coronal geometry.

The right panel shows the iron-abundance measurements. In this case, the systematic uncertainty introduced by differences in coronal geometry, disk electron density, and model choice is comparable to the quoted statistical uncertainty.

\section{Conclusion}
\label{sec:conclusion}

In this work, we have focused on quantifying the systematic uncertainties that arise from different coronal geometries in spectral fitting. By comparing four widely used models— the power-law emissivity profile, broken power-law emissivity profile, the lamppost corona, and the disk-like corona. We have shown how these model assumptions influence key spectral parameters, including the black hole spin, inclination, ionization and the iron abundance.

Our analysis reveals that the model-dependent range of spectral parameters is significant. Specifically, the black hole spin and inclination can be inferred to be relatively low, with values as low as $\sim0.7$ and $\sim30^\circ$ respectively, depending on the chosen coronal geometry. This highlights the importance of carefully considering the assumed coronal model when interpreting black hole spin and inclination angle measurements, as different geometries can lead to different conclusions.

While each model provides a statistically reasonable fit to the data, our results demonstrate that the disk-like corona model yields more physically consistent results. This model, supported by recent polarization measurements \citep{Henric2022}, suggests a lower inclination angle ($\sim 30^\circ$), consistent with the binary orbital inclination ($27.1^\circ \pm 0.8^\circ$; \citealt{Orosz2011}), and resolves the unphysical negative outer emissivity index found in the power-law models. Additionally, the disk-like model constrains the electron density to $n_e \approx 10^{20}\, \text{cm}^{-3}$, which aligns with theoretical expectations for the inner disk regions.

The current implementation of the disk-like corona model has limitations: the fixed outer radius of the disk-shaped coronal region and the ignored processes such as Compton scattering of reflection photons or re-interaction of inverse Compton photons with the coronal region may limit its realism. The slight statistical inferiority, coupled with these simplifications, indicates that spectral data alone are insufficient to definitively favour a disk-like geometry over the lamppost scenario.

However, it is essential to recognize that the primary aim of this work is not to definitively determine the "correct" coronal geometry, but rather to assess how different assumptions about the corona's shape and size affect the interpretation of the spectral data. The disk-like model provides an appealing physical interpretation, but other models, such as the lamppost or broken power-law geometries, may still be valid in certain contexts, depending on additional observational constraints.

To refine our understanding of the geometry and its effects on spectral parameters, future work should incorporate more detailed models and consider multi-messenger approaches, combining both spectroscopic and polarimetric data to further constrain the physical properties of the corona in sources like \src.

\label{conclusion}

\appendix

\section{Additional plots}

\begin{figure}[htbp]
    \hspace{-1cm}
    \includegraphics[width=1.15\linewidth]{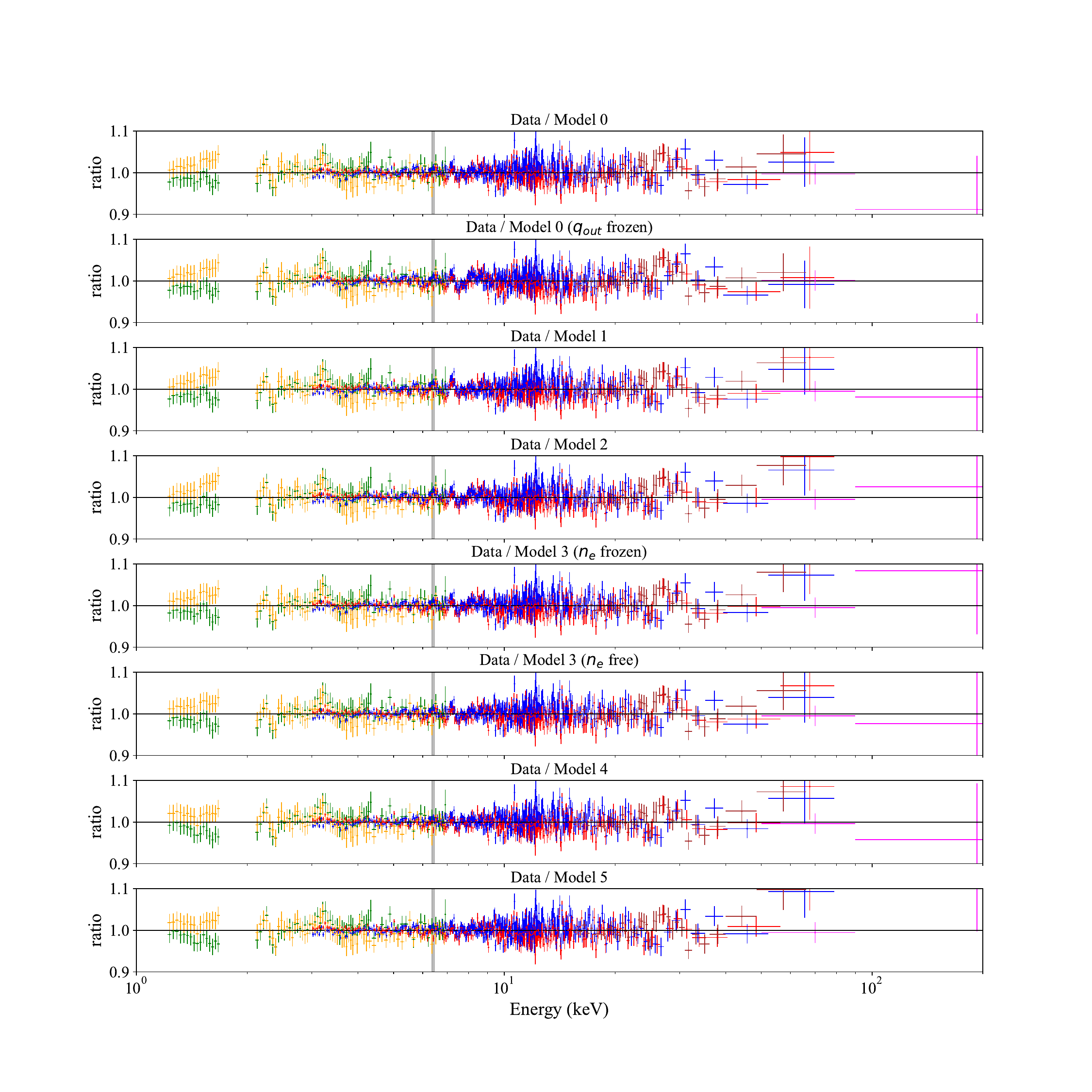}
    \caption{Ratio plots (data/model) for all model combinations are provided for the 1-200 keV energy range, with corresponding titles. The colored error bars—red, blue, green, orange, brown, and magenta—represent data from {\em NuSTAR}'s FPMA, FPMB, and {\em Suzaku}'s XIS0, XIS1, PIN, and GSO detectors, respectively. A thick gray line marks the iron K$\alpha$ line at 6.4 keV, with a 0.1 keV width.}
    \label{fig:ratio-full}
\end{figure}

Figure~\ref{fig:ratio-full} presents the ratio (data/model) plots over the 1-200 keV energy range for all model combinations discussed in this paper, with appropriate titles included. For further details, please refer to the caption.

\section{MCMC plots}
\label{sec:mcmc-plots}

\begin{figure}[htbp]
    \centering
    \includegraphics[width=1.\linewidth]{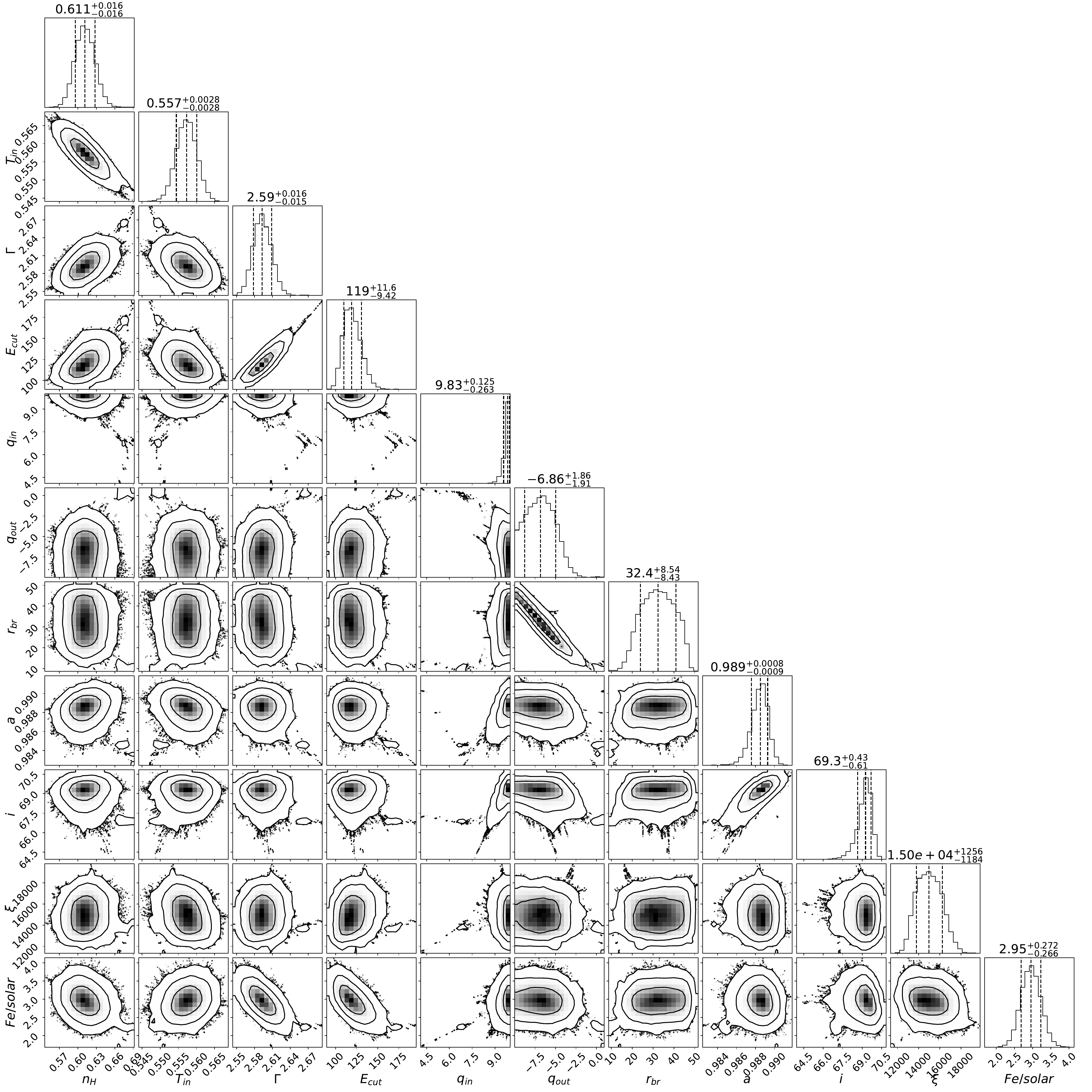}
    \caption{Corner plot of MCMC result of Model 0.}
    \label{fig:mcmc-relconv}
\end{figure}

The results of the MCMC simulations for Models 0, 3, 4, and 5 are presented sequentially below. Figure~\ref{fig:mcmc-relconv} illustrates the MCMC analysis of Model 0 and provides contour plots depicting the relationships between pairs of the following parameters: $n_H$ (interstellar hydrogen column density), $T_{\rm in}$ (accretion disk temperature), $Fe/solar$ (iron abundance in solar units), $\xi$ (ionization parameter), $a$ (black hole spin), $i$ (inclination angle of the accretion disk), $q_{\rm in}$ (inner emissivity index), $q_{\rm out}$ (outer emissivity index), $r_{\rm br}$ (breaking radius), $E_{\rm cut}$ (high-energy cutoff), and $\Gamma$ (photon index).

\begin{figure}[htbp]
    \centering
    \includegraphics[width=1.\linewidth]{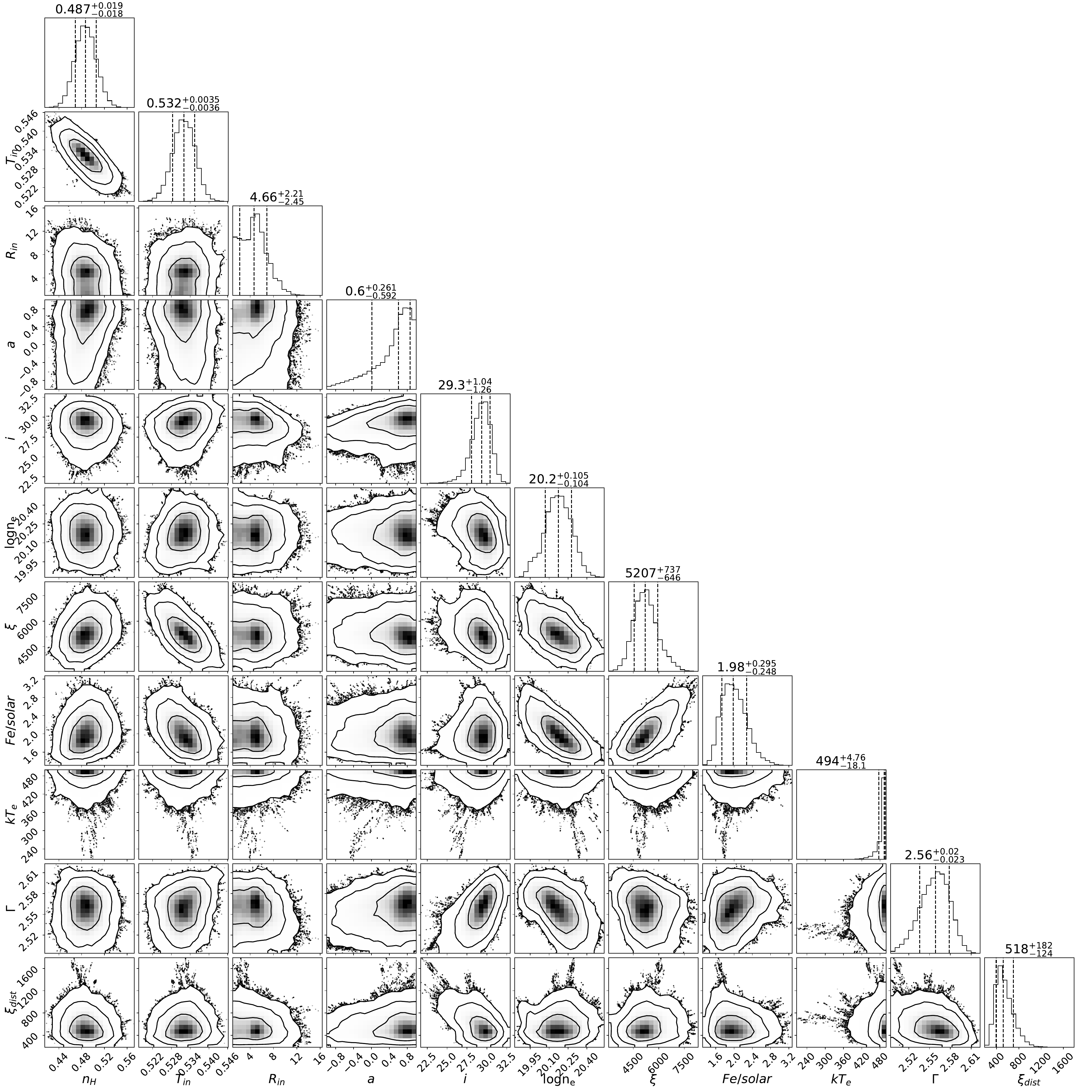}
    \caption{Corner plot of MCMC result of Model 3. Here $\xi$ and $\xi_{\rm dist}$ are the ionisation parameters of main and distant reflectors, respectively.}
    \label{fig:mcmc-relconvdisk}
\end{figure}

Similarly, Figure~\ref{fig:mcmc-relconvdisk} displays the MCMC analysis of Model 3, incorporating a free electron density, and provides contour plots illustrating the relationships between pairs of the following parameters: $n_H$ (interstellar hydrogen column density), $T_{\rm in}$ (accretion disk temperature), $Fe/solar$ (iron abundance in solar units), $\xi$ (ionization of the main reflector), $\xi_{\rm dist}$ (ionization of the distant reflector), $a$ (black hole spin), $i$ (inclination angle of the accretion disk), $R_{\rm in}$ (inner radius of the coronal disk), $kT_{\rm e}$ (coronal electron temperature), $\log n_{\rm e}$ (electron density in logarithmic scale), and $\Gamma$ (photon index).

\begin{figure}[htbp]
    \centering
    \includegraphics[width=1.\linewidth]{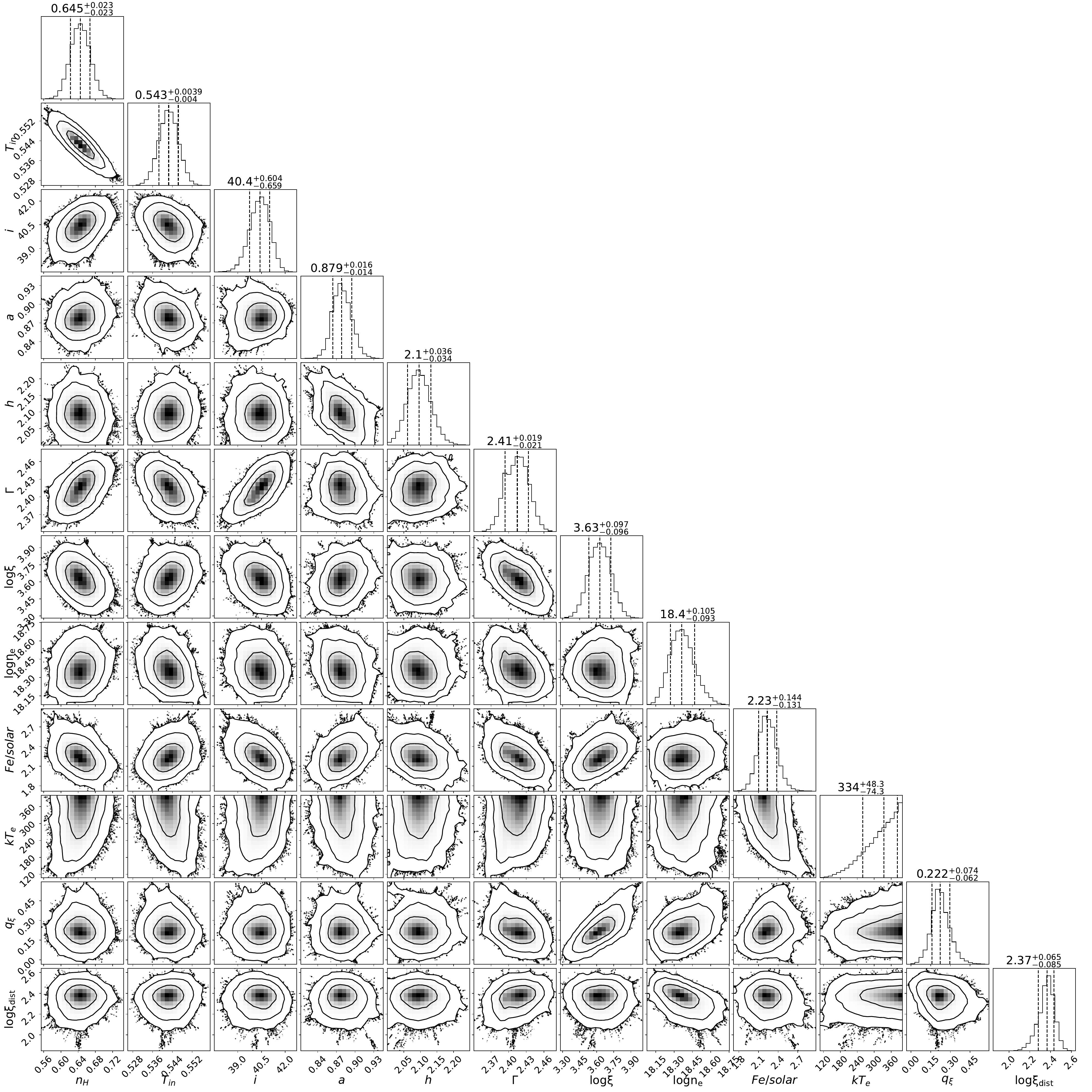}
    \caption{Corner plot of MCMC result of Model 4. Here $\log\xi$ and $\log\xi_{\rm dist}$ are ionisation parameters of main and distant reflectors on a logarithmic scale, respectively.}
    \label{fig:mcmc-relxillCp}
\end{figure}

\begin{figure}[htbp]
    \centering
    \includegraphics[width=1.\linewidth]{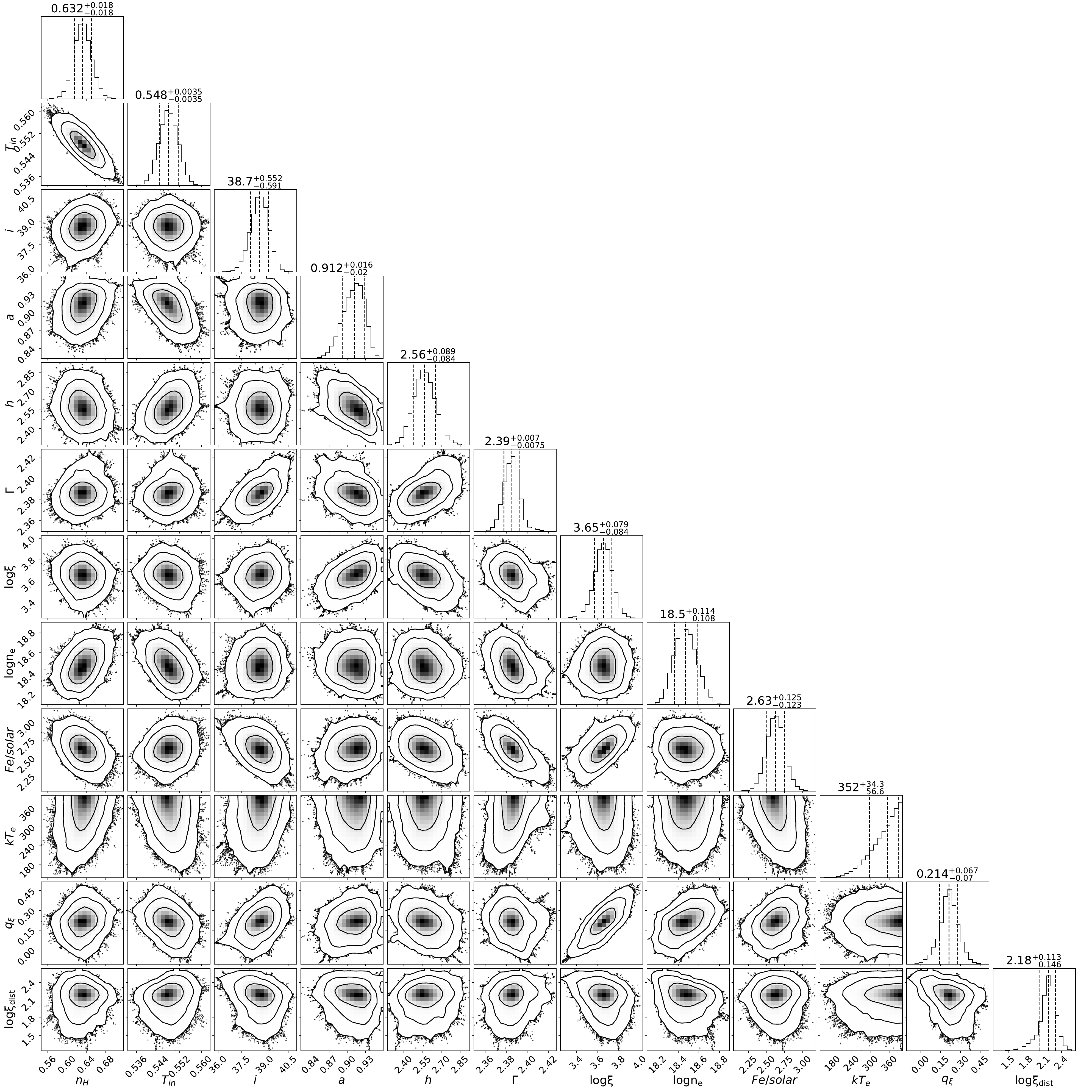}
    \caption{Corner plot of MCMC result of Model 5. Here $\log\xi$ and $\log\xi_{\rm dist}$ are ionisation parameters of main and distant reflectors on a logarithmic scale, respectively.}
    \label{fig:mcmc-relxillCp-boost}
\end{figure}

Figures~\ref{fig:mcmc-relxillCp} and \ref{fig:mcmc-relxillCp-boost} present the MCMC analyses for Models 4 and 5, respectively, and display contour plots depicting the relationships between pairs of the following parameters: $n_H$ (interstellar hydrogen column density), $T_{\rm in}$ (accretion disk temperature), $Fe/solar$ (iron abundance in solar units), $\log\xi$ (ionization at the inner edge of the accretion disk, in logarithmic scale), $q_{\xi}$ (ionization profile index), $\log\xi_{\rm dist}$ (ionization of the distant reflector, in logarithmic scale), $a$ (black hole spin), $i$ (inclination angle of the accretion disk), $h$ (height of the corona), $kT_{\rm e}$ (electron temperature in the corona), $\log n_{\rm e}$ (electron density in logarithmic scale), and $\Gamma$ (photon index).

{\bf Acknowledgments --}
This work was supported by the National Natural Science Foundation of China (NSFC), Grant No.~W2531002, and the Fudan-Warwick Joint Seed Fund {\it Probing the Inner Accretion Flow of Black Holes: A Synergy Between Warwick and Fudan}, Grant No.~JMH6282518.

\bibliography{references}
\bibliographystyle{aasjournalv7}

\end{document}

%% file: t-models.tex
\begin{table*}[htpb]
\hspace{-1cm}
\begin{adjustbox}{width=\textwidth}
\begin{tabular}{ll}
\tableline
\\[-3ex]
\tableline
Model & XSPEC model 
\\ \tableline
0 & ${\tt constant \cdot 
xstar 
\cdot TBabs \cdot 
( diskbb + cutoffpl + relconv \times  reflionx ) }$ \\
1 & ${\tt constant \cdot 
xstar 
\cdot  TBabs \cdot ( diskbb + cutoffpl + relconv \times reflionx + reflionx )}$ \\
2 & ${\tt constant \cdot xstar \cdot  
TBabs \cdot (diskbb + nthComp+  relconvdisk\_nk \times 
reflionx)}$ \\
3 & ${\tt constant \cdot  xstar \cdot  TBabs \cdot 
( diskbb + nthComp + relconvdisk\_nk \times  reflionx + reflionx)}$ \\
4 & ${\tt constant \cdot xstar \cdot  TBabs \cdot ( diskbb + nthComp+ relxilllpCp  + xillverCp)}$ \\
5 & ${\tt constant \cdot xstar \cdot  TBabs \cdot ( diskbb + relxilllpCp  + xillverCp)}$ 
\\ \tableline
\end{tabular}
\end{adjustbox}
\caption{List of all model combinations used in this work.}
\label{tab:models}
\end{table*}

%% file: t-relconv.tex
\begin{table*}
\hspace{-1cm}
\begin{adjustbox}{width=\textwidth}
\renewcommand{\arraystretch}{1.05}
\begin{tabular}{lllll}
\tableline
\\[-3ex]
\tableline
\textbf{Parameter} & \textbf{Unit/Description} & \textbf{Model 0} & \textbf{Model 0 ($q_{\rm out}$=3)} & \textbf{Model 1} 
\\ \tableline
Simple ionized absorber &  &  & & \\ 
$N_{\rm H}$ & $10^{22} \, \rm{cm}^{-2}$ & $3.8^{+0.1}_{-0.4}$ & $2.3^{+0.2}_{-0.1}$ & $3.5^{+0.2}_{-0.2}$ \\ 
$\rm{log} \xi$ & $\xi$ in erg cm $s^{-1}$ & $>4.9$ & $4.38^{+0.04}_{-0.03}$ & $4.69^{+0.09}_{-0.05}$ 
\\ \tableline
Interstellar absorbtion & & &  & \\ 
$N_{\rm H}$ & $10^{22} \, \rm{cm}^{-2}$ & $0.61^{+0.02}_{-0.01}$ & $0.63^{+0.01}_{-0.01}$ & $0.59^{+0.01}_{-0.01}$ 
\\ \tableline
Disk blackbody & &  &  & \\ 
$kT_{\rm in}$ & keV  & $0.558^{+0.003}_{-0.001}$ & $0.553^{+0.002}_{-0.003}$ & $0.560^{+0.004}_{-0.003}$ \\ 
norm & Normalization $(\times 10^{4})$ & $2.11^{+0.03}_{-0.03}$ & $2.11^{+0.09}_{-0.04}$ & $2.13^{+0.06}_{-0.10}$ 
\\ \tableline
Cutoff power-law & &  &  & \\ 
$\Gamma$ & Photon index & $2.590^{+0.004}_{-0.003}$ & $2.650^{+0.006}_{-0.008}$ & $2.54^{+0.03}_{-0.03}$ \\ 
$E_{\rm cut}$ & keV  & $115^{+4}_{-3}$ & $172^{+24}_{-10}$ & $96^{+12}_{-9}$ \\ 
norm & Normalization & $6.18^{+0.08}_{-0.13}$ & $6.6^{+0.1}_{-0.2}$ & $5.6^{+0.3}_{-0.2}$ 
\\ \tableline
Relativistic blurring (\texttt{relconv}) & &  &  & \\ 
$q_{\rm in}$ & Emissivity index & $>9.3$ & $>9.8$ & $>9.6$ \\ 
$q_{\rm out}$ & Emissivity index & $-6.96^{+0.07}_{-0.04}$ & $3\footnote{Fixed.}$ & $3^{\rm{a}}$ \\ 
$r_{\rm break}$ & Breaking radius ($r_{g}$)  & $31^{+5}_{-5}$ & $97$ & $12$ \\ 
${a_*}$ & BH spin & $0.989^{+0.001}_{-0.002}$ & $0.93^{+0.02}_{-0.01}$ & $0.986^{+0.001}_{-0.004}$ \\ 
$i$ & Inclination (deg) & $69.8^{+0.2}_{-1.1}$ & $60.6^{+0.5}_{-0.9}$ & $69.1^{+0.6}_{-1.8}$ 
\\ \tableline
Reflection component (\texttt{reflionx}) & &  &  & \\ 
$\xi$ & $10^{3}$ erg cm s $^{-1}$ & $14.7^{+0.3}_{-0.6}$ & $>19.3$ & $>18.8$ \\ 
Fe/solar & Abundance & $3.0^{+0.2}_{-0.2}$ & $2.0^{+0.14}_{-0.06}$ & $3.7^{+0.3}_{-0.4}$ \\ 
norm & Normalization $(\times 10^{-6})$ & $7.1^{+0.1}_{-1.2}$ & $6.5^{+0.3}_{-0.1}$ & $4.1^{+0.4}_{-0.2}$ 
\\ \tableline
Distant reflector (\texttt{reflionx}) & &  &  & \\ 
$\xi$ & $10^{3}$ erg cm s $^{-1}$ & $----$ & $----$ & $13^{+4}_{-3}$ \\ 
norm & Normalization $(\times 10^{-6})$ & $----$ & $----$ & $1.3^{+0.4}_{-0.4}$ \\ \tableline
Cross-normalisation constants (relative to FPMA) & &  &  & \\ 
$C_{FPMB}$ & ...  & $1.001^{+0.001}_{-0.001}$ & $1.001^{+0.001}_{-0.001}$ & $1.001^{+0.001}_{-0.001}$ \\ 
$C_{XIS0}$ & ...  & $1.077^{+0.005}_{-0.005}$ & $1.076^{+0.005}_{-0.005}$ & $1.076^{+0.005}_{-0.005}$ \\ 
$C_{XIS1}$ & ... & $1.036^{+0.005}_{-0.005}$ & $1.035^{+0.004}_{-0.002}$ & $1.036^{+0.005}_{-0.005}$ \\ 
$C_{PIN}$ & ... & $1.202^{+0.006}_{-0.006}$ & $1.203^{+0.006}_{-0.006}$ & $1.201^{+0.005}_{-0.006}$ \\ 
$C_{GSO}$ & ... & $1.23^{+0.05}_{-0.05}$ & $1.18^{+0.04}_{-0.05}$ & $1.27^{+0.05}_{-0.05}$ 
\\ \tableline
Goodness of fit & &  &  & \\ 
$\chi^2/\nu$ & ... & 1285.3/1095 & 1422.7/1096 & 1328.9/1094 
\\ \tableline
\end{tabular}
\end{adjustbox}
\caption{Best fit parameter values for Model 0 and Model 1 with 90\% confidence range.}
\label{relconv2}
\end{table*}

%% file: t-relconvdisk.tex
\begin{table*}
\hspace{-1cm}
\begin{adjustbox}{width=\textwidth}
\renewcommand{\arraystretch}{1.1}
\begin{tabular}{lllll}
\tableline
\\[-3ex]
\tableline
\textbf{Parameter} & \textbf{Unit/Description} & \textbf{Model 2 } & \textbf{Model 3 ($n_{\rm e}$ frozen)} & \textbf{Model 3 ($n_{\rm e}$ free)}  
\\ \tableline
Simple ionized absorber & &  &  & \\ 
$N_{\rm H}$ & $10^{22} \, \rm{cm}^{-2}$ & $2.3^{+0.2}_{-0.2}$ & $4.2^{+0.4}_{-0.2}$ & $3.8^{+0.7}_{-0.3}$ \\ 
$\rm{log} \xi$ & $\xi$ in erg cm $\rm{s}^{-1}$ & $4.48^{+0.10}_{-0.03}$ & $4.84^{+0.15}_{-0.04}$ & $>4.75$ 
\\ \tableline
Interstellar absorbtion &  &  &  & \\ 
$N_{\rm H}$ & $10^{22} \, \rm{cm}^{-2}$  & $0.41^{+0.02}_{-0.02}$ & $0.47^{+0.01}_{-0.01}$ & $0.50^{+0.03}_{-0.02}$ 
\\ \tableline
Disk blackbody & &  &  & \\ 
$kT_{\rm in}$ & keV   & $0.55^{+0.0003}_{-0.0033}$ & $0.54^{+0.004}_{-0.002}$ & $0.532^{+0.0041}_{-0.0003}$ \\ 
norm & Normalization $(\times 10^{4})$  & $2.4^{+0.1}_{-0.1}$ & $2.72^{+0.07}_{-0.09}$ & $2.9^{+0.2}_{-0.2}$ 
\\ \tableline
Comptonization component & &  &  & \\ 
norm & Normalization  & $0.9^{+0.3}_{-0.2}$ & $1.3^{+0.1}_{-0.4}$ & $1.2^{+0.4}_{-0.1}$ 
\\ \tableline
Relativistic blurring (\texttt{relconvdisk\_nk}) &  &  &  & \\ 
h & Height of corona ($r_g$) & $>6$ & $<5$ & $<4.5$ \\ 
$R_{\rm in}$ & Inner radius of corona ($r_g$)  & $<5$ & $<10$ & $<6$ \\ 
$R_{\rm out}$ & Outer radius of corona ($r_g$) & $24\footnote{Fixed.}$ & $24^{\rm{a}}$ & $24^{\rm{a}}$ \\ 
${a_*}$ & BH spin  & $0.82^{+0.08}_{-0.38}$ & $>-0.8$ & $0.73^{+0.21}_{-0.07}$ \\ 
$i$ & Inclination (deg)  & $27^{+1.4}_{-0.5}$ & $31.2^{+1.8}_{-0.9}$ & $30.6^{+0.5}_{-2.7}$ 
\\ \tableline
Reflection component (\texttt{reflionx}) &  &  &  & \\ 
$\xi$ & $10^{3}$ erg cm s $^{-1}$ & $14^{+1}_{-2}$ & $14^{+1}_{-3}$ & $5^{+1}_{-2}$ \\ 
Fe/solar & Abundance & $>4.94$ & $3.8^{+0.3}_{-0.4}$ & $2.0^{+0.6}_{-0.3}$ \\ 
$kT_{\rm e}$ & keV & $>384$ & $>332$ & $>185$ \\ 
$\Gamma$  & Photon index  & $2.55^{+0.03}_{-0.02}$ & $2.65^{+0.04}_{-0.01}$ & $2.56^{+0.05}_{-0.09}$ \\
$\log n_{\rm e}$ & $n_{\rm e}$ in cm$^{-3}$ & $15^{\rm{a}}$ & $15^{\rm{a}}$ & $20.2^{+0.3}_{-1.0}$ \\
norm & Normalization  & $48^{+4}_{-1}$ & $47^{+2}_{-1}$ & $11.3^{+0.03}_{-1.52}$ 
\\ \tableline
Distant reflector (\texttt{reflionx}) & &  &  & \\ 
$\xi$ & erg cm $\rm{s}^{-1}$  & $----$ & $168^{+24}_{-15}$ & $484^{+347}_{-166}$ \\ 
norm & Normalization  & $----$ & $79^{+23}_{-17}$ & $2.5^{+1.8}_{-0.6}$ 
\\ \tableline
Cross-normalisation constants (relative to FPMA) & &  &  & \\ 
$C_{FPMB}$ & ...  & $1.001^{+0.001}_{-0.001}$ & $1.001^{+0.001}_{-0.001}$ & $1.001^{+0.001}_{-0.001}$ \\ 
$C_{XIS0}$ & ...   & $1.076^{+0.005}_{-0.005}$ & $1.075^{+0.005}_{-0.005}$ & $1.075^{+0.005}_{-0.005}$ \\ 
$C_{XIS1}$ & ...   & $1.036^{+0.005}_{-0.005}$ & $1.035^{+0.005}_{-0.005}$ & $1.034^{+0.005}_{-0.005}$ \\ 
$C_{PIN}$ & ...   & $1.201^{+0.006}_{-0.006}$ & $1.201^{+0.003}_{-0.006}$ & $1.201^{+0.006}_{-0.006}$ \\ 
$C_{GSO}$ & ...   & $1.30^{+0.05}_{-0.05}$ & $1.30^{+0.05}_{-0.05}$ & $1.26^{+0.05}_{-0.05}$ 
\\ \tableline
Goodness of fit & &  &  & \\ 
$\chi^2/\nu$  & ...   & 1359.3/1096 & 1307.4/1094 & 1296.5/1093 
\\ \tableline
\end{tabular}
\end{adjustbox}
\caption{Best fit parameter values for Model 2 and Model 3 with 90\% confidence range.}
\label{tab:relconvdisk}
\end{table*}

%% file: t-relxilllpCp.tex
\begin{table*}
\hspace{-1cm}
\begin{adjustbox}{width=\textwidth}
\renewcommand{\arraystretch}{1.05}
\begin{tabular}{llll}
\tableline
\\[-3ex]
\tableline
\textbf{Parameter} & \textbf{Unit/Description} & \textbf{Model 4} & \textbf{Model 5}  
\\ \tableline
Simple ionized absorber & &  & \\ 
$N_{\rm H}$ & $10^{22} \, \rm{cm}^{-2}$ & $3.64^{+0.09}_{-0.09}$ & $3.3^{+0.3}_{-0.3}$ \\ 
$\rm{log} \xi$ & $\xi$ in erg cm $s^{-1}$ & $>4.95$ & $>4.8$ 
\\ \tableline
Interstellar absorbtion &  &  & \\ 
$N_{\rm H}$ & $10^{22} \, \rm{cm}^{-2}$  & $0.653^{+0.008}_{-0.008}$ & $0.64^{+0.02}_{-0.02}$ 
\\ \tableline
Disk blackbody &  &  & \\ 
$kT_{\rm in}$ & keV    & $0.541^{+0.0003}_{-0.0003}$ & $0.547^{+0.004}_{-0.004}$ \\ 
norm & Normalization $(\times 10^{4})$  & $2.386^{+0.008}_{-0.008}$ & $2.3^{+0.1}_{-0.1}$ 
\\ \tableline
Comptonization component & &  & \\ 
norm & Normalization & $0.634^{+0.006}_{-0.006}$ & $----$ 
\\ \tableline
Relativistic reflection model (\texttt{relxilllpCp}) &  &  & \\ 
$i$ & Inclination (deg) & $40.6^{+0.2}_{-0.2}$ & $38.9^{+0.6}_{-0.7}$ \\ 
${a_*}$ & BH spin   & $0.875^{+0.002}_{-0.002}$ & $0.92^{+0.02}_{-0.03}$ \\ 
h& Height of corona ($r_g$)  & $<2.1$ & $2.6^{+0.3}_{-0.3}$ \\ 
$\Gamma$  & Photon index   & $2.424^{+0.0008}_{-0.0008}$ & $2.388^{+0.008}_{-0.007}$ \\ 
$\rm{log} \xi$ & $\xi$ in erg cm $s^{-1}$ & $3.605^{+0.005}_{-0.005}$ & $3.65^{+0.08}_{-0.09}$ \\ 
$\log n_{\rm e}$ & $n_{\rm e}$ in cm$^{-3}$ & $18.341^{+0.006}_{-0.006}$ & $18.4^{+0.2}_{-0.1}$ \\ 
Fe/solar & Abundance & $2.14^{+0.02}_{-0.02}$ & $2.6^{+0.2}_{-0.1}$ \\ 
$kT_{\rm e}$ & keV & $>346$ & $>284$ \\ 
iongrad_index & Power-law ionization profile index  & $0.218^{+0.004}_{-0.004}$ & $0.21^{+0.08}_{-0.06}$ \\ 
refl_frac & Reflection fraction &	-1\footnote{Fixed.} & 1$^{\rm{a}}$	\\
iongrad_type & Flag for power-law ionisation profile \footnote{\texttt{iongrad_type}=1 means power-law ionisation gradient ($\rm{log} \xi \propto r^{-\text{iongrad\_index}}$) and constant density.} & 1 & 1 \\
switch_returnrad & Flag for returning radiation & 1 & 1 \\
switch_reflfrac_boost & Flag for self-consistent comptonization & 0 & 1 \\
norm  & Normalization & $1.68^{+2.618}_{-0.003}$ & $0.20^{+0.06}_{-0.03}$ 
\\ \tableline
Distant reflector (\texttt{xillverCp}) & &  & \\ 
$\rm{log} \xi$ & $\xi$ in erg cm $\rm{s}^{-1}$ & $2.39^{+0.02}_{-0.02}$ & $2.2^{+0.2}_{-0.2}$ \\ 
norm  & Normalization ($\times 10^{-3}$) & $8.9^{+0.3}_{-0.3}$ & $4.6^{+1.8}_{-0.9}$ 
\\ \tableline
Cross-normalisation constants (relative to FPMA) & &  & \\ 
$C_{FPMB}$ & ...  & $1.001^{+0.001}_{-0.001}$ & $1.001^{+0.001}_{-0.001}$ \\ 
$C_{XIS0}$ & ...  & $1.076^{+0.004}_{-0.004}$ & $1.076^{+0.005}_{-0.005}$ \\ 
$C_{XIS1}$ & ...  & $1.036^{+0.004}_{-0.004}$ & $1.036^{+0.005}_{-0.005}$ \\ 
$C_{PIN}$ & ...  & $1.202^{+0.005}_{-0.005}$ & $1.202^{+0.006}_{-0.006}$ \\ 
$C_{GSO}$ & ...  & $1.28^{+0.05}_{-0.05}$ & $1.33^{+0.05}_{-0.05}$ 
\\ \tableline
Goodness of fit &  &  & \\ 
$\chi^2/\nu$ & ...  & 1240.3/1093 & 1249.3/1094 
\\ \tableline
\end{tabular}
\end{adjustbox}
\caption{Best fit parameter values for Model 4 and Model 5 with 90\% confidence range.}
\label{tab:relxilllpCp2}
\end{table*}